# *Indication of anomalous heat energy production in a reactor device containing hydrogen loaded nickel powder.*

Giuseppe Levi
Bologna University, Bologna, Italy

Evelyn Foschi
Bologna, Italy

Torbjörn Hartman, Bo Höistad, Roland Pettersson and Lars Tegnér
Uppsala University, Uppsala, Sweden

Hanno Essén
Royal Institute of Technology, Stockholm, Sweden

**ABSTRACT**

An experimental investigation of possible anomalous heat production in a special type of reactor tube named *E-Cat HT* is carried out. The reactor tube is charged with a small amount of hydrogen loaded nickel powder plus some additives. The reaction is primarily initiated by heat from resistor coils inside the reactor tube. Measurement of the produced heat was performed with high-resolution thermal imaging cameras, recording data every second from the hot reactor tube. The measurements of electrical power input were performed with a large bandwidth three-phase power analyzer. Data were collected in two experimental runs lasting 96 and 116 hours, respectively. An anomalous heat production was indicated in both experiments.
The 116-hour experiment also included a calibration of the experimental set-up without the active charge present in the *E-Cat HT*. In this case, no extra heat was generated beyond the expected heat from the electric input.
Computed volumetric and gravimetric energy densities were found to be far above those of any known chemical source. Even by the most conservative assumptions as to the errors in the measurements, the result is still one order of magnitude greater than conventional energy sources.

**INTRODUCTION**

Andrea Rossi claims to have invented an apparatus that can produce much more energy per unit weight of fuel than can be obtained from known chemical processes. His invention is referred to as an energy catalyzer named *E-Cat HT*, where *HT* stands for high temperature. The original idea behind Rossi's invention goes back to experiments done in the nineties by Sergio Focardi at Bologna University and collaborators, in which they claimed to have observed an anomalous heat production in a hydrogen-loaded nickel rod [1-2]. Later, an experiment [3] was carried out by S. Focardi and A. Rossi using an apparatus with a sealed container holding nickel powder plus unknown additives pressurized with hydrogen gas. When the container was heated, substantial heat was produced in excess of the input heat. They speculated that a "low energy nuclear reaction" had taken place in order to explain the large amount of excess heat. The *E-Cat HT* – a further, high temperature development of the original apparatus which has also undergone many construction changes in the last two years – is the latest product manufactured by Leonardo Corporation: it is a device allegedly capable of producing heat from some type of reaction the origin of which is unknown.
As in the original E-Cat, the reaction is fueled by a mixture of nickel, hydrogen, and a catalyst, which is kept as an industrial trade secret. The charge sets off the production of thermal energy after having been activated by heat produced by a set of resistor coils located inside the reactor. Once operating temperature is reached, it is possible to control the reaction by regulating the power to the coils.
The scope of the present work is to make an independent test of the *E-Cat HT* reactor under controlled conditions and with high precision instrumentation. It should be emphasized that the measurement must be performed with high accuracy and reliability, so that any possible excess heat production can be established beyond any doubt, as no known processes exist which can explain any abundant heat production in the E-Cat reactor.

The present report describes the results obtained from evaluations of the operation of the *E-Cat HT* in two test runs. The first test experiment, lasting 96 hours (from Dec. 13th 2012, to Dec. 17th 2012), was carried out by the two first authors of this paper, Levi and Foschi, while the second experiment, lasting for 116 hours (from March 18th 2013, to March 23rd 2013), was carried out by all authors. Both experiments were performed on the premises of EFA Srl, Via del Commercio 34-36, Ferrara (Italy).

The tests held in December 2012 and March 2013 are in fact subsequent to a previous attempt in November 2012 to make accurate measurements on a similar model of the *E-Cat HT* on the same premises. In that experiment the device was destroyed in the course of the experimental run, when the steel cylinder containing the active charge overheated and melted. The partial data gathered before the failure, however, yielded interesting results which warranted further in-depth investigation in future tests. Although the run was not successful as far as obtaining complete data is concerned, it was fruitful in that it demonstrated a huge production of excess heat, which however could not be quantified. The device used had similar, but not identical, features to those of the *E-Cat HT* used in the December and March runs.

Besides some minor geometrical differences, in the *E-Cat HT* used for the November test the charge in the inner cylinder was not evenly distributed, but concentrated in two distinct locations along the central axis. In addition, the primer resistor coils were run at about 1 kW, which might be the cause of the ensuing device failure. For these reasons, a more prudent reactor design was chosen for the test held in December and March, by distributing the charge evenly along its container cylinder, and limiting the power input to the reactor to 360 W.

Since the test in November shows some interesting features, we shall describe some of the results from this test in some detail before discussing, in the subsequent sections, the results from the December and March runs. Figures 1 and 2 refer to the November test, and show, respectively, the device while in operation, and a laptop computer capturing data from a thermographic camera focused on it. An Optris IR camera monitored surface temperature trends, and yielded results of approximately 860°C in the hottest areas.

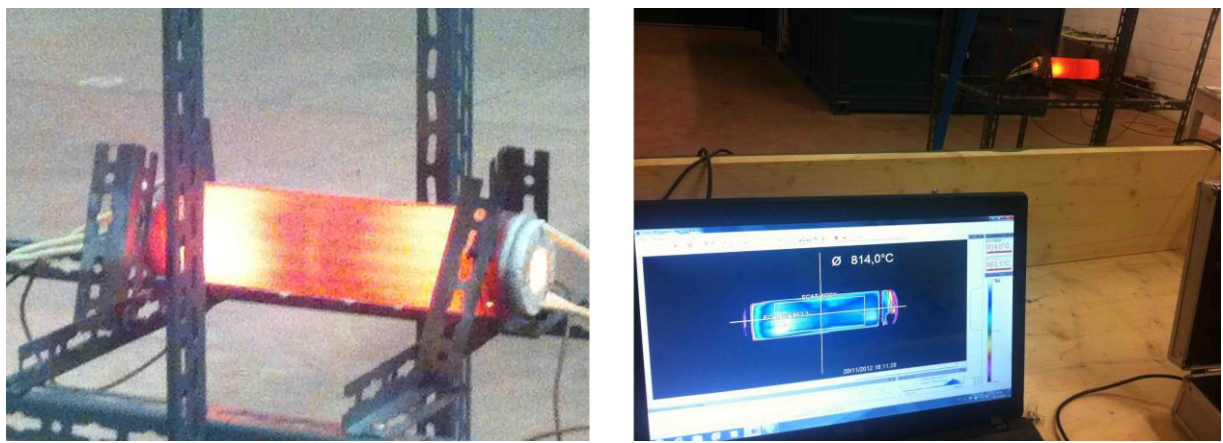

*Figs. 1-2. Two images from the test performed on Nov. 20th 2012. Here, the activation of the charge (distributed laterally in the reactor) is especially obvious. The darker lines in the photograph are actually the shadows of the resistor coils, which yield only a minimal part of the total thermal power. The performance of this device was such that the reactor was destroyed, melting the internal steel cylinder and the surrounding ceramic layers. The long-term trials analyzed in the present report were purposely performed at a lower temperatures for safety reasons.*

Fig. 3 shows a thermal video frame from the IR camera: the temperature of 859°C refers to Area 2 (delimited by the "cross hairs"), whereas the average temperature recorded for the body of the device, relevant to the rectangle indicated as Area 1, is 793°C.

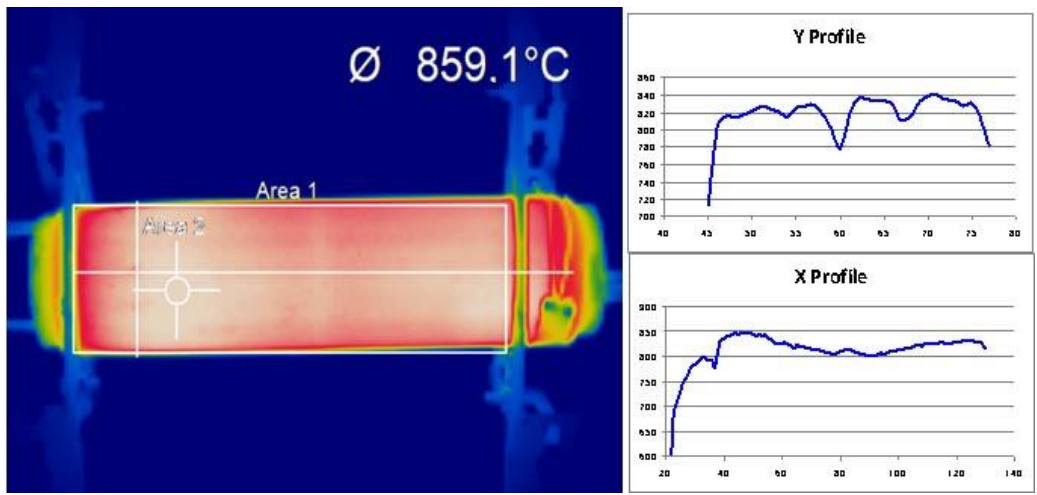

*Fig. 3. Thermal image of the November test device. The temperature of 859°C refers to the area within the circle of the mark (Area 2). The graphs on right show the temperature trends along the horizontal line traversing the device (X Profile), and along the vertical line on the left of the image (Y Profile).*

Graphs on the right side of fig. 3 show the temperature distribution monitored along the two visible lines in the image: the X Profile refers to the horizontal line traversing the whole device, the Y Profile shows the temperature along the vertical line located on the left side of the thermal image. This latter distribution allows one to reach some interesting conclusions.

If one relates the length of the vertical line (32 pixels) to the diameter of the device (11 cm), one may infer that each pixel in the image corresponds to a length of approximately 0.34 cm on the device (with some approximation, due to the fact that the thermal image is a two-dimensional projection of a cylindrical object). The thermal image shows a series of stripe-like, darker horizontal lines, which are confirmed by the five temperature dips in the Y Profile. This means that, in the device image, a darker line appears every 6.4 pixels approximately, corresponding to 2.2 cm on the device itself. As mentioned previously, the *E-Cat HT* needs resistor coils in order to work; these are set horizontally, parallel to and equidistant from the cylinder axis, and extend throughout the whole length of the device. By dividing the circumference of the base of the cylinder by the number of coils, one may infer that the 16 resistor coils in this device were laid out at a distance of 2.17 cm. one from the other. And, by comparing the distance between darker stripes and the distance between coils, one may reach the conclusion that the lower temperatures picked up by the thermal camera nicely match the areas overlying the resistor coils. In other words, the temperature dips visible in the diagram are actually shadows of the resistor coils, projected on the camera lens by a source of energy located further inside the device, and of higher intensity as compared to the energy emitted by the coils themselves.

## PART 1: THE DECEMBER TEST

**Device and experimental set-up**
The *E-Cat HT*-type device in this experiment was a cylinder having a silicon nitride ceramic outer shell, 33 cm in length, and 10 cm in diameter. A second cylinder made of a different ceramic material (corundum) was located within the shell, and housed three delta-connected spiral-wire resistor coils. Resistors were laid out horizontally, parallel to and equidistant from the cylinder axis, and were as long as the cylinder itself. They were fed by a TRIAC power regulator device which interrupted each phase periodically, in order to modulate power input with an industrial trade secret waveform. This procedure, needed to properly activate the *E-Cat HT* charge, had no bearing whatsoever on the power consumption of the device, which remained constant throughout the test. The most important element of the *E-Cat HT* was lodged inside the structure. It consisted of an AISI 310 steel cylinder, 3 mm thick and 33 mm in diameter, housing the powder charges. Two AISI 316 steel cone-shaped caps were hot-hammered in the cylinder, sealing it hermetically. Cap adherence was obtained by exploiting the higher thermal expansion coefficient of AISI 316 with respect to AISI 310 steel.

Finally, the outermost shell was coated by a special aeronautical-industry grade black paint capable of withstanding temperatures up to 1200°C.

It was not possible to evaluate the weight of the internal steel cylinder or of the caps because the *E-Cat HT* was already running when the test began. Weighing operations were therefore performed on another perfectly similar device present on the premises, comparing a cap-sealed cylinder containing the active charge with another identical cylinder, empty and without caps. The difference in weight obtained is 0.236 kg: this is therefore to be assigned to the charge loaded into the *E-Cat HT* and to the weight (not subtracted in the present test) of the two metal caps.

In the course of the test, the *E-Cat HT* was placed on a metal frame and allowed to freely radiate to the surrounding air. The contact points between the device and the frame were reduced to the minimum necessary for mechanical stability; room temperature was constantly measured by means of a heat probe, and averaged 15.7°C (= 289 K).

The instruments used to acquire experimental data were at all times active for the entire 96 hours of the test, and consisted of an IR thermography camera to measure the *E-Cat HT*'s surface temperature, and a wide band-pass power quality monitor measuring the electrical quantities on each of the three phases, to record the power absorbed by the resistor coils.

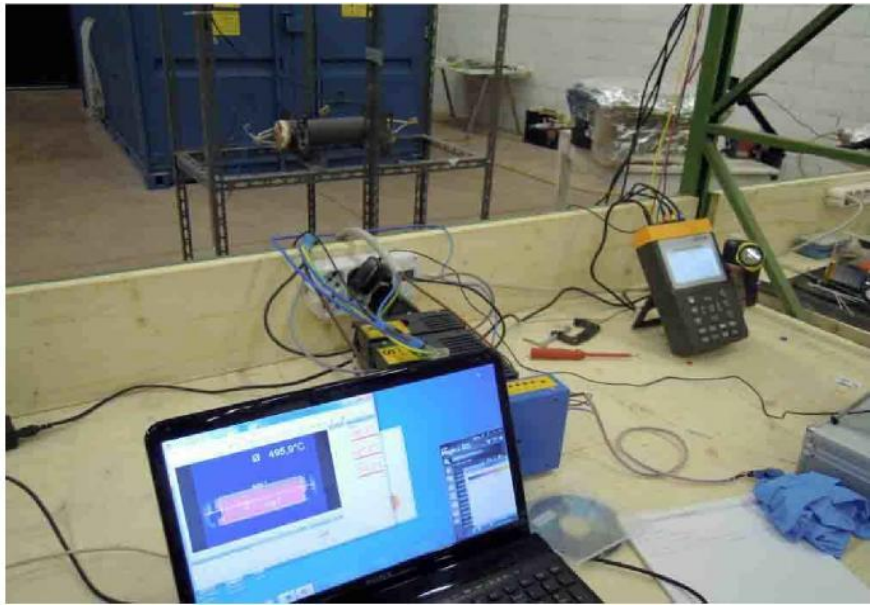

*Fig. 4. Instrumentation set-up for measurements. Foreground: thermal image capture. On right: instrument used for electrical measurements. Background: the* E-Cat HT *on its support frame; the IR camera is not visible here.*

The thermal camera used was an Optris PI 160 Thermal Imager with 30° × 23° lens, and UFPA 160 × 120 pixel sensors. The camera spectral interval is from 7.5 to 13 μm, with a precision of 2% of measured value. The camera was fastened to the frame of the *E-Cat HT*, and positioned about 70 cm from the device, with lens facing the lower half of the cylinder. All imaging was thus taken from below the apparatus, in order not to damage the lens from the heat transferred by rising convective air currents. This choice, however, had a negative impact on the measurements: the presence of the two metal props on the stationary image shot by the camera introduced a degree of uncertainty in the measurements, as will be explained in detail below.Camera capture rate was set at 1 Hz, and the image, visualized on a laptop display, was open to analysis throughout the course of the test.

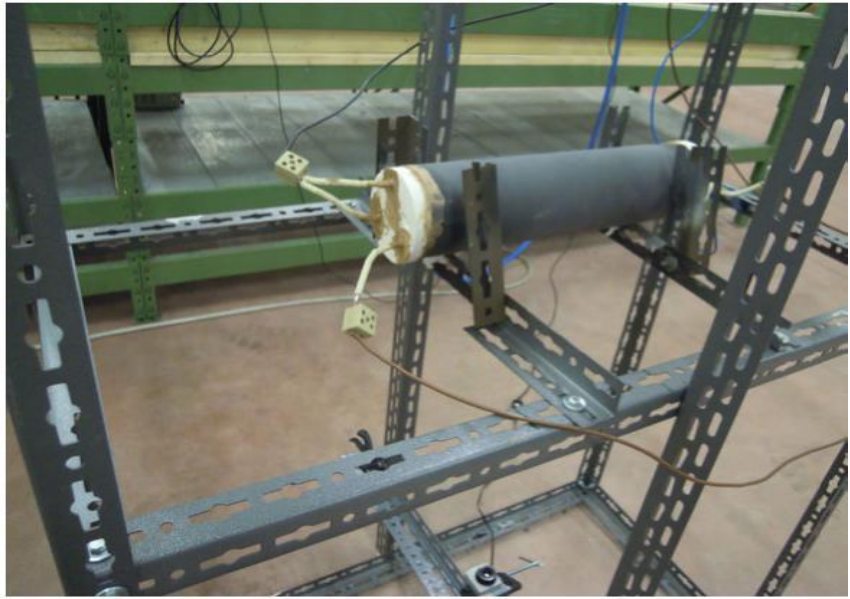

*Fig. 5. E-Cat HT on support frame. The power cables to the internal resistor coils are visible, as well as the IR camera in the lower part of the photograph.*

Electrical measurements were performed by a PCE-830 Power and Harmonics Analyzer by PCE Instruments with a nominal accuracy of 1%. This instrument continuously monitors on an LCD display the values of instantaneous electrical power (active, reactive, and apparent) supplied to the resistor coils, as well as energy consumption expressed in kWh.

Of these parameters, only the last one was of interest for the purposes of the test, which was designed to evaluate the ratio of thermal energy produced by the *E-Cat HT* to electrical power consumption for the number of hours subject to evaluation. The instrument was connected directly to the *E-Cat HT* cables by means of three clamp ammeters, and three probes for voltage measurement.

A wristwatch was placed next to the wattmeter, and a video camera was set up on a tripod and focused on both objects: at one frame per second, the entire sequence of minutes and power consumption were filmed and recorded for the 96-hour duration of the test.

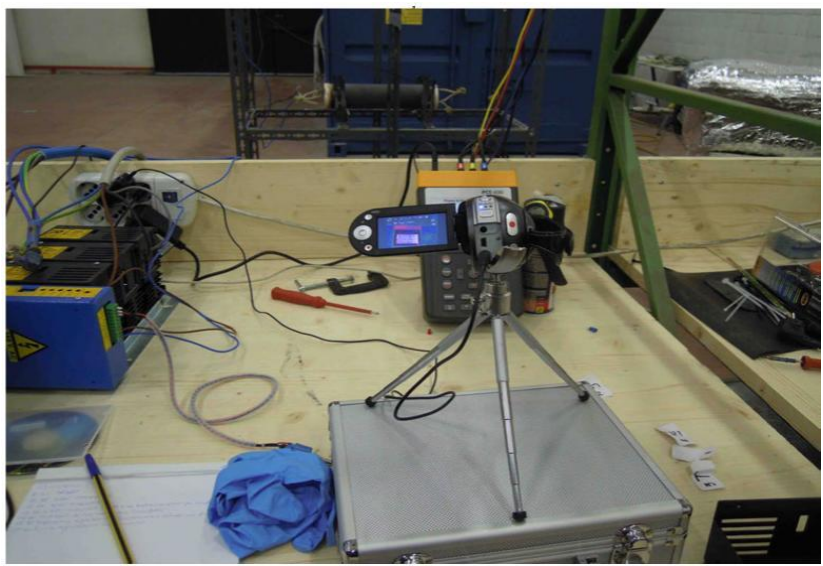

*Fig. 6. The video camera on its tripod framing the display used for electrical measurements (PCE-830), and recording at 1 frame p.s. for the whole duration of the test.*

Besides the set-up required for the measurements, instruments necessary for detecting possible radioactive emissions were also placed in the vicinity of the *E-Cat HT*. These measurements are essential for safety certification of the device, and were performed by David Bianchini. The full report of the methods and results of these measurements is available on demand. A partial quotation follows:

"*It was decided to use two different wide-spectrum and high-sensitivity photon detectors: the first detector was chosen for the purpose of measuring in the spatial surroundings any rate variation of ambient dose equivalent H * (10) [...], the second detector was chosen for measuring and recording CPM (counts per minute) rate variations in a specific position [...]. With respect to instrumental and ambient background, the measurements performed do not reveal significant differences either in H\*(10) or CPM ascribable to the E-Cat prototype*".

**Data analysis**

Upon conclusion of the test, the recordings from the video camera were examined. By reading the images reproducing the PCE-830's LCD display at regular intervals, it was possible to make a note of the number of kWh absorbed by the resistor coils. Subsequently, the *E-Cat HT's* average hourly power consumption was calculated, and determined to be = 360 W.

As far as the evaluation of the energy produced by the *E-Cat HT* is concerned, two dominant components must be taken into account, the first being emission by radiation, the second the dispersion of heat to the environment by means of convection.

Heat transfer by conduction was deemed to be negligible, due to of the minimal surface of contact (not more than a few $mm^2$) between the device and its supports, and to the fiberglass insulation material placed at the contact points. This material, however, partially obscured the image of the *E-Cat HT's* surface.

Energy emitted by radiation was calculated by means of Stefan-Boltzmann's formula, which allows to evaluate the heat emitted by a body when its surface temperature is known.

Surface temperature was measured by analyzing the images acquired by the IR camera, after dividing the images into multiple areas, and extracting the average temperature value associated to each area.

Conservatively, surface emissivity during measurements was set to 1, i.e. the temperature values recorded are consequently lower than real, as will be explained below.

The calculation of energy loss by convection from objects of cylindrical shape placed in air has been presented many times in academic papers that address issues related to heat transfer (see for example [4,5]). It was therefore possible to estimate the amount of heat transferred by the *ECat HT* to the surrounding air in the course of the test.

The thermal performance of the *E-Cat HT* was finally obtained as the ratio between the total energy emitted by the device and the energy dissipated by its resistor coils.

**Calculating the power emitted by radiation**

Planck's Law expresses how the monochromatic emissive power of a black body varies as a function of its absolute temperature and wavelength; integrating this over the whole spectrum of frequencies, one obtains the total emissive power (per unit area) of a black body, through what is known as Stefan-Boltzmann's Law:

$M=\sigma T^4$ [$W/m^2$]     (1)

where σ indicates Stefan-Boltzmann's constant, equal to $5.67 \cdot 10^{-8}$ [$W/m^2K^4$].

In the case of real surfaces, one must also take emissivity (ε) into account. ε expresses the ratio between the energy emitted from the real surface, and that which would be emitted by a black body having the same temperature. The formula then becomes:

$M=\varepsilon\sigma T^4$ [$W/m^2$]     (2)

where ε may vary between 0 and 1, the latter value being the one assumed for a black body. As it was not possible to measure the emissivity of the coating used in this analysis, it was decided to conservatively assume a value of ε = 1, thereby considering the *E-Cat HT* as equivalent to a black body. This value was then input in the thermal imagery software, which allows the user to modify some of the parameters, such as ambient temperature and emissivity, even after having completed the recordings. The camera software then uses the new settings to recalculate the temperature values assigned to the recorded images. It was therefore possible to determine the *E-Cat HT*'s emitted thermal power on the basis of surface temperature values that were never overestimated with respect to actual ones.

The veracity of this statement may be proven by an example where we see what happens when one assigns a value lower than 1 to ε: in fig. 7, the thermal image of the *E-Cat HT* has been divided into 40 areas. Emissivity has been set to = 1 everywhere, except in two areas (Nos. 18 and 20), where it is set to 0.8 and 0.95, respectively. The temperature which the IR camera assigns to the two areas is 564.1°C and 511.7°C, respectively – these values being much higher than those of the adjacent areas.

It is therefore obvious that by assigning a value of 1 to ε in to every area, we are in fact performing a conservative measurement: this is a necessary precaution, given the lack of information on the real emissivity value of the *E-Cat HT*.

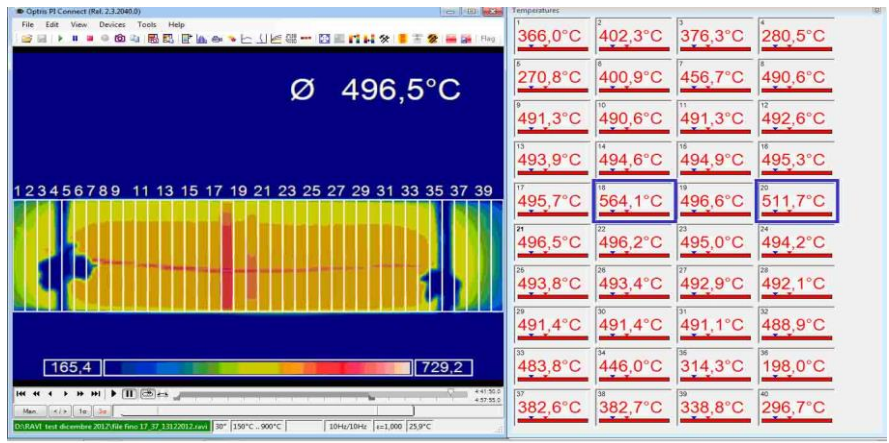

*Fig. 7. This image exemplifies the effect of emissivity on the determination of temperatures. The E-Cat HT has been divided in 40 areas; in areas 18 and 20 emissivity has been set to = 0.8 and 0.95, respectively, whereas in all the remaining areas ε has been set to = 1. Measured temperatures appear to be higher in areas 18 and 20 with respect to those recorded in the other areas. If the lower values for ε were extended to all areas, this would lead to a higher estimate of irradiated energy density. For our calculations, therefore – in view of the fact that the effective value of ε was not available for our test, and that it was felt desirable to avoid any arbitrary source of overestimation – ε was left set to = 1 in all areas.*

One must keep in mind that the thermal camera does not measure an object's temperature directly: with the help of input optics, radiation emitted from the object is focused onto an infrared detector which generates a corresponding electrical signal. Digital signal processing then transforms the signal into an output value proportional to the object temperature. Finally, the temperature result is shown on the camera display. The camera software derives the temperature of objects by means of an algorithm which takes several parameters and corrective factors into account, e.g. user settings for emissivity and detector temperature, taken automatically by a sensor on the lower part of the camera itself.

Moreover, every Optris camera-and-optics set has its own calibration file supplied by the manufacturer (Ref. [6]).

The image provided by the camera shows only the lower part of the *E-Cat HT*: as no other IR cameras were available, the same temperatures measured there were held good for the upper half of the device as well, and were used for subsequent calculations. We realize that convection has a

- 7 -

different effect on the top of the object compared to the bottom of it. Therefore, the temperature values by means of the frame setup chosen for positioning the camera should be the ones least affected by convective dispersion. This choice, however, leads to a heavy penalty in the calculation of the average surface temperature of the *E-Cat HT*: as a matter of fact, in the frames associated with the setup, the shadows of the two metal struts, and of the insulating materials placed under the device to support it, are clearly visible. These two blacked-out areas negatively distort the calculation of the surface temperature and prevent a proper view of the underlying emitting surfaces.

To overcome this problem, it was decided to divide the entire image of the IR camera into a progressively greater number of areas, for which average temperature values for the entire duration (96 hrs) of the test were calculated. Subsequently, these values were raised to the 4th power, and then averaged together to obtain a single value to be assigned to the *E-Cat HT*. By this process, the blacked-out areas are actually considered as pertaining to the surface of the *E-Cat HT*, thereby underestimating the energy emitted. It was decided to proceed in this manner in order to obtain a lower limit for emitted energy based solely on collected data, without making arbitrary assumptions that might have led to errors by overestimation.

The image obtained from the IR camera covers an area of $160 \times 41$ pixels and was progressively divided into 10, 20 and 40 areas, following the following criterion: in the first case, 10 areas of $16 \times 41$ pixels; in the second, 20 areas of $8 \times 41$ pixels; finally, in the third, 40 areas of $4 \times 41$ pixels.

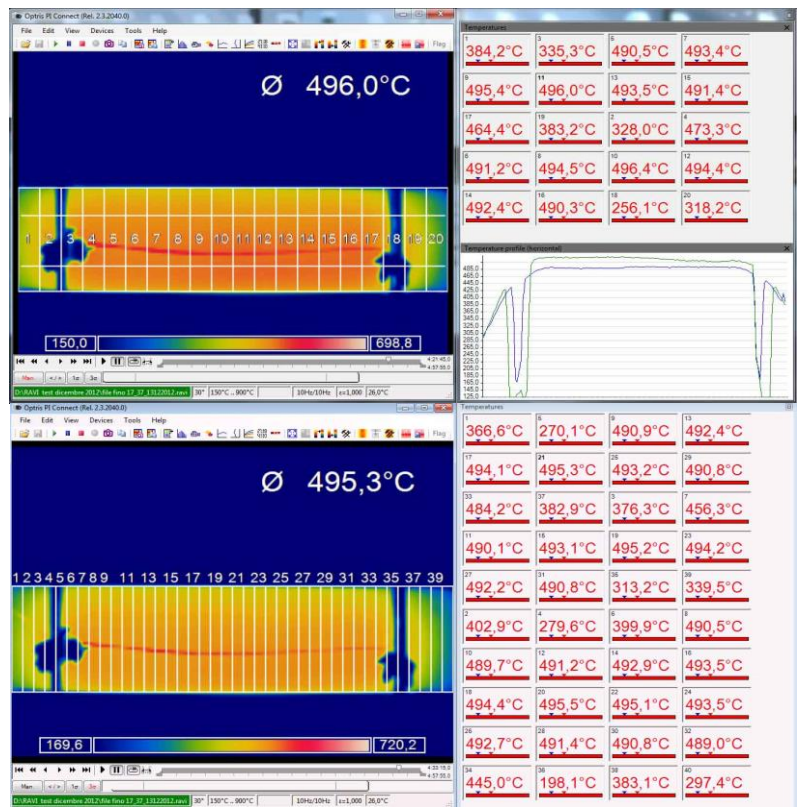

*Fig. 8: The area occupied by the* E-Cat HT *was first divided into 10 parts, then into 20, and finally into 40 parts. For each of these the average temperature was measured.*
*Above: 20-area division of the* E-Cat HT *image. The thermal profiles on the right refer to the two horizontal lines on the camera image: in the profiles, the two temperature dips corresponding to the metal struts are clearly visible.*
*Below: 40-area division of the* E-Cat HT *image. The red horizontal line crossing the image is due to a small crack in the ceramic outer surface caused by mechanical stress probably due to a previous thermal shock.*

For each area, as well as for the entire duration of the video footage, a time diagram of the average temperature trend was extracted; data was then saved to Excel worksheets, from which the averages were extracted.

The temperatures thus obtained, expressed in Kelvin for each area, are presented in the following three tables.

| Area 1 | Area 2 | Area 3 | Area 4 | Area 5 | Area 6 | Area 7 | Area 8 | Area 9 | Area 10 |
|---|---|---|---|---|---|---|---|---|---|
| 628.8 K | 623.8 K | 665.1 K | 754.3 K | 759.3 K | 761.8 K | 761.2 K | 759.0 K | 756.4 K | 624.8 K |

*Table 1. Division into 10 areas.*

By averaging these 10 values, one obtains a temperature, associable to the *E-Cat HT*, of 709 K.

| Area 1 | Area 2 | Area 3 | Area 4 | Area 5 | Area 6 | Area 7 | Area 8 | Area 9 | Area 10 |
|---|---|---|---|---|---|---|---|---|---|
| 660.9 K | 596.4 K | 599.0 K | 738.9 K | 757.0 K | 757.9 K | 760.1 K | 761.1 K | 762.0 K | 763.0 K |
| **Area 11** | **Area 12** | **Area 13** | **Area 14** | **Area 15** | **Area 16** | **Area 17** | **Area 18** | **Area 19** | **Area 20** |
| 762.7 K | 761.3 K | 760.5 K | 760.0 K | 758.7 K | 757.3 K | 732.0 K | 521.8 K | 650.5 K | 592.8 K |

*Table 2. Division into 20 areas.*

By averaging these 20 values, one obtains an assignable temperature for the *E-Cat HT* of 710.7 K.

| Area 1 | Area 2 | Area 3 | Area 4 | Area 5 | Area 6 | Area 7 | Area 8 | Area 9 | Area 10 |
|---|---|---|---|---|---|---|---|---|---|
| 641.6 K | 670.7 K | 644.5 K | 546.0 K | 535.3 K | 667.4 K | 724.0 K | 758.4 K | 758.8 K | 757.9 K |
| **Area 11** | **Area 12** | **Area 13** | **Area 14** | **Area 15** | **Area 16** | **Area 17** | **Area 18** | **Area 19** | **Area 20** |
| 758.5 K | 759.7 K | 761.1 K | 762.0 K | 762.4 K | 762.9 K | 763.4 K | 763.6 K | 764.5 K | 764.9 K |
| **Area 21** | **Area 22** | **Area 23** | **Area 24** | **Area 25** | **Area 26** | **Area 27** | **Area 28** | **Area 29** | **Area 30** |
| 764.7 K | 764.5 K | 763.6 K | 763.0 K | 762.9 K | 762.5 K | 762.0 K | 761.3 K | 760.7 K | 760.9 K |
| **Area 31** | **Area 32** | **Area 33** | **Area 34** | **Area 35** | **Area 36** | **Area 37** | **Area 38** | **Area 39** | **Area 40** |
| 760.7 K | 758.6 K | 753.3 K | 713.2 K | 581.4 K | 463.5 K | 652.2 K | 652.6 K | 608.1 K | 564.4 K |

*Table 3. Division into 40 areas.*

By averaging these 40 values, one may assign to the *E-Cat HT* a temperature of 711.5 K.

The comparison between the different subdivisions into areas shows that the average temperature depends only slightly upon the choice of subdivision, and actually tends to increase, because the areas near the blacked-out ones are treated more effectively.

With reference to the third case above, one may calculate thermal power emitted by the *ECat HT* by first considering the average of the fourth power of the temperature of each area. One gets the following value:

$$(T^4)_{average} = 2.74 \cdot 10^{11} \, [K^4] \tag{3}$$

Emitted thermal power (*E*) may be easily obtained by multiplying the Stefan-Boltzmann formula by area of the *E-Cat HT* :

$$Area_{E\text{-}Cat} = 2\pi RL = 1036 \cdot 10^{-4} \, [m^2] \tag{4}$$

where:
R = radius of the *E-Cat HT*, equal to 0.05 [m]
L = length of the *E-Cat HT*, equal to 0.33 [m].

$$E = (5.67 \cdot 10^{-8})(2.74 \cdot 10^{11})(1036 \cdot 10^{-4}) = 1609 \ [W] \tag{5}$$

In calculating the total area of the *E-Cat HT*, the area of the two bases was omitted, their surface being:

$$Area_{\text{E-Cat Bases}} = 2(\pi R^2) = 157 \cdot 10^{-4} \ [m^2] \tag{6}$$

This choice was motivated by the fact that for these parts of the cylinder, which are not framed by the IR camera, any estimate of irradiated energy would have been highly conjectural. We therefore preferred not to include this factor in calculating *E*, thereby underestimating radiative thermal power emitted by the *E-Cat HT*.

Emitted thermal power (*E*), apart from minute variations, remains constant throughout the measurement, as may be seen in the Plot 1 below, showing the measured radiative power vs time in hours. Power production is almost constant with an average of 1609.4 W.

To this power we must subtract the thermal power due to room temperature. On the basis of an average of 15.7°C over 96 hours, we get:

$$E\_room = (5.67 \cdot 10^{-8})(289)^4 (1036 \cdot 10^{-4}) = 41 \ [W] \tag{7}$$

So the final value is:

$$E - E\_room = 1609 - 41 = 1568 \ [W] \tag{8}$$

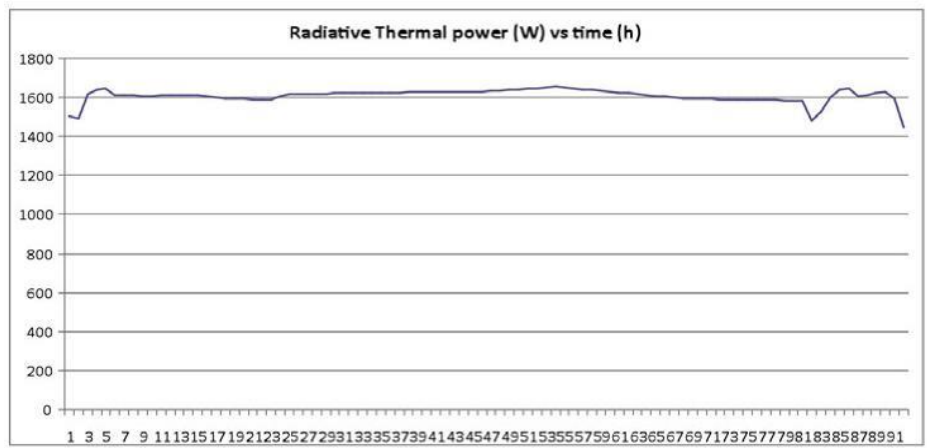

*Plot 1. Emitted thermal power vs time. Power production is almost constant with an average of 1609.4 W.*

Note that the image reproduced by the IR camera is actually the projection of a cylindrical object on a two-dimensional plane. Consequently, the lines of sight between the camera and the cylinder radius vary between 0 and 90 degrees. In the latter case, which refers to the lateral parts of the *E-Cat HT* with respect to camera position, and thereby to the edges of the thermal image, the recorded temperatures may be significantly lower than effective ones. However, the division into rectangles adopted by us in order to calculate the average temperatures, comprises these edges (see fig. 8), which will therefore appear to be colder than they actually are due to the IR camera's angle of view. Once again, we opted to take a conservative stance, underestimating temperatures where the effective value was not easily assessable.

**Calculating power emitted by convection**

Let us consider a fluid temperature $Tf$ lapping against a surface having area A and temperature T. Heat Q transferred in unit time by convection between the surface and the fluid may be expressed by Newton's relation:

$$Q = hA\ (T-Tf) = hA\ \Delta T\ [W] \tag{9}$$

where h is defined as the heat exchange coefficient [W/m$^2$ K].
When the value of h is known, it is possible to evaluate the heat flow; thus, determining h constitutes the fundamental problem of thermal convection. Convection coefficient h is not a thermo-physical property of the fluid, but a parameter the value of which depends on all the variables that influence heat exchange by convection:

$h = f\ (\rho,\ Cp,\ \mu,\ \beta g,\ k,\ T\text{-}Tf,\ D)$, where the meaning of the symbols is as follows:

$\rho$ = fluid density [kg/m³]
$Cp$: specific heat capacity at constant pressure [J/kgK]
$\mu$: viscosity [kg/ms]
$\beta g$: product of the coefficient of thermal expansion by gravity acceleration [m/s² K]
k: coefficient of thermal conductivity [W/mK]
$T\text{-}Tf = \Delta T$: temperature difference between surface and fluid [K]
D: linear dimension; in our case, diameter [m].

The value of h may be obtained, for those instances involving the more common geometries and those fluids of greater practical interest, through the use of expressions resulting from experimental tests quoted in mainstream heat engineering literature. With reference to these texts [4, 5], we see that, in the case of a cylinder with a diameter less than 20 cm immersed in air at a temperature close to 294 K, the value of h may be had through the following expression:

$$h = C''\ (T\text{-}T_f)^n\ D^{3n-1} \tag{10}$$

C'' and n are two constants the value of which may be obtained if one knows the interval within which the product between the Grashof number $Gr$ and the Prandtl number $Pr$ falls. These dimensionless numbers are defined as follows:

$$G_r = \beta g\ (T\text{-}T_f)\ D^3\ \rho^2/\mu^2; \qquad P_r = C_p\mu/k \tag{11}$$

$G_r$ represents the ratio between the inertia forces of buoyancy and friction forces squared, while $P_r$ represents the ratio between the readiness of the fluid to carry momentum and its readiness to transport heat.
For a wide range of temperatures one can say that:

$$k^4\ (\beta g\rho^2\ C_p\ /\mu k) = 36.0 \tag{12}$$

For the *E-Cat HT* average temperature value derived above, we get an average temperature between device and air equal to:

$$(T + Tf)/2 = (711.5 + 289)/2 = 500.2\ [K] \tag{13}$$

Once this value is known, one can first of all derive the relevant coefficient of thermal conductivity k. With the aid of Table 4, which holds good for air, the value of k obtained for this temperature is equal to 0.041 [W/mK].

| k [W/mK] | T [K] |
|---|---|
| 0.0164 | 173 |
| 0.0242 | 273 |
| 0.0317 | 373 |
| 0.0391 | 473 |
| 0.0459 | 573 |

*Table 4. The extreme temperature values given constitute the experimental range. For extrapolation to other temperatures, it is suggested that the data given be plotted as log k vs log T (see reference [4]).*

From (12) we have:

$(\beta g \rho^2 C_p)/(\mu k) = 36.0/(0.041)^4 = 1.27 \cdot 10^7$ (14)

From the definitions of $G_r$ and $P_r$ we get:

$G_r P_r = ((\beta g \rho^2 C_p)/(\mu k))(T-T_f) L^3 = (1.27 \cdot 10^7)(711.5-289)(0.1)^3 = 5.36 \cdot 10^6$ (15)

Now we may consult Table 5 for the two constants we are searching for:

| *GrPr* | n | C'' |
|---|---|---|
| $10^{-5} - 10^{-3}$ | 0.04 | - |
| $10^{-3} - 1.0$ | 0.10 | - |
| $1.0 - 10^4$ | 0.20 | - |
| $10^4 - 10^9$ | 0.25 | 1.32 |
| $>10^9$ | 0.33 | 1.24 |

*Table 5. Values are referred to a horizontal cylinder with a diameter less than 0.2 m (see ref.[4]).*

One may then deduce:

C'' = 1.32, n = 0.25

(10) then becomes:

$h = (1.32)(711.5-289)^{0.25}(0.1)^{3*0.25-1} = 10.64$ [W/m² K] (16)

Substituting (16) in (9) we obtain the power emitted by convection:

$Q = (10.64)(1036 \cdot 10^{-4})(711.5-289) = 466$ [W] (17)

### *E-Cat HT* performance calculation

At this point all that remains to be done, in order to get the performance (COP) of the *E-Cat HT*, is to add the radiated power to the power dispersed by convection, and relate the result to the power supplied to the heating elements. Conservatively, we can associate to these values a percentage error of 10%, in order to comprise various sources of uncertainty: those relevant to the consumption measurements of the *E-Cat HT*, those inherent in the limited range of frequencies upon which the IR cameras operate, and those linked to the calculation of average temperatures.

From (8) and from (17) we have:

$$1568 + 466 = (2034 \pm 203) \text{ [W]} \tag{18}$$

$$COP = 2034/360 = 5.6 \pm 0.8 \tag{19}$$

assuming a 10% error in the powers. Plot 2 shows produced vs. consumed energy. Radiated energy is actually measured energy; total energy also takes into account the evaluation of natural convection. Data are fit with a linear function, and COP is obtained by the slope.

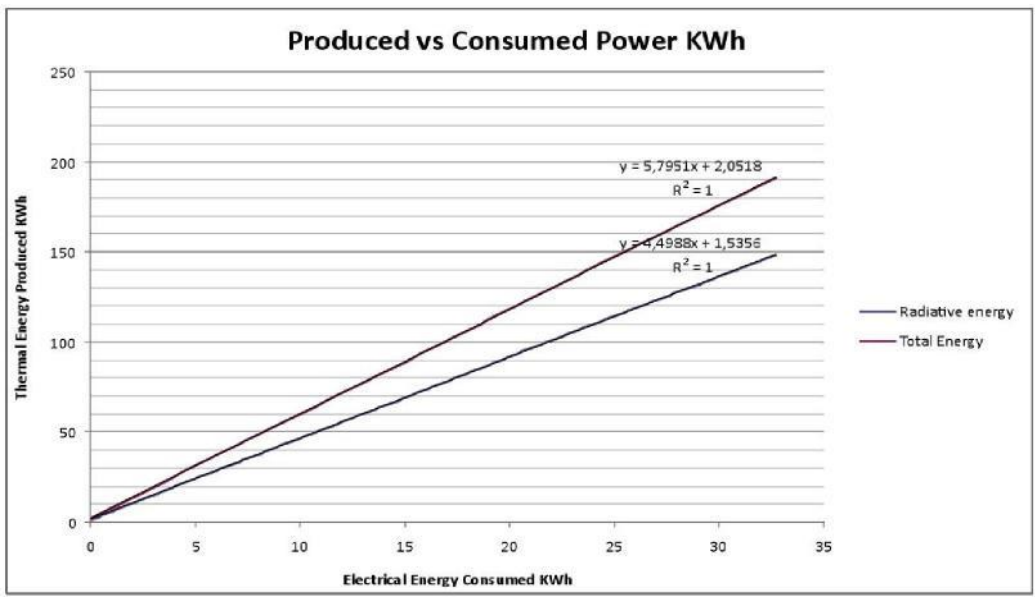

*Plot 2. Thermal energy produced (kWh) versus electrical energy consumed (kWh).*

**Ragone Chart**

As we have seen, the weight of the active charge of the *E-Cat HT* plus the weight of the two metal caps sealing the inner cylinder is equal to 0.236 kg.
If we assign this value to the charge powders, we shall be overestimating the weight of the charge; thus, our calculation of the values of power density and the density of thermal energy may be regarded as a lower limit.

For power density we have:

$$(2034-360)/0.236 = (7093 \pm 709) \text{ [W/kg]} \tag{20}$$

Thermal energy density is obtained by multiplying (20) by the number of test hours:

$$7093 \cdot 96 = 680949 \text{ [Wh/kg]} \sim (6.81 \pm 0.7) \cdot 10^5 \text{ [Wh/kg]} \tag{21}$$

Figure 9 shows the "Ragone plot of energy storage", a typical diagram in which specific energy is represented as a function on a logarithmic scale of the specific power of the various energy storage technologies [Ref. 7]. Power density and thermal energy density found for the *E-Cat HT* place this device outside of the area occupied by any known conventional energy source in the Ragone chart. Given the deliberately conservative choices made in performing the measurement, we can reasonably state that the *E-Cat HT* is a non-conventional source of energy which lies between conventional chemical sources of energy and nuclear ones.

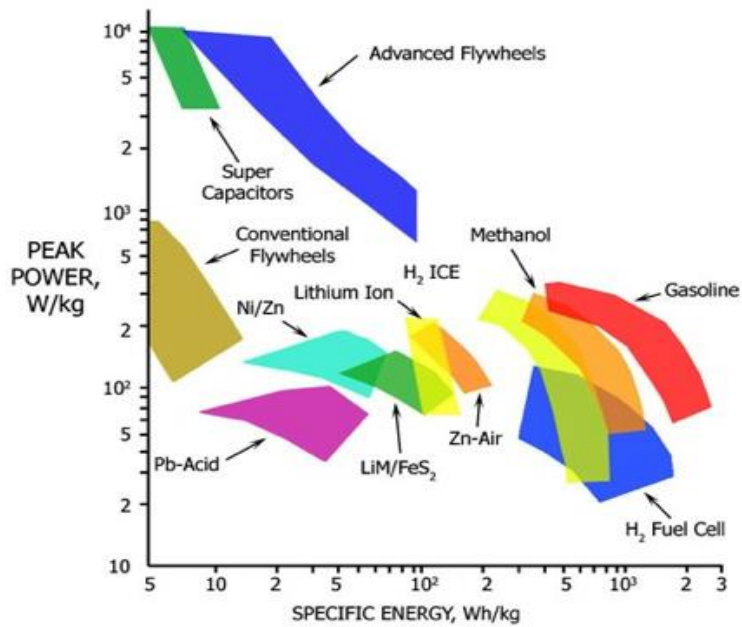

*Fig. 9. "Ragone plot of energy storage"[Ref. 7]. The plot shows specific gravimetric energy and power densities relevant to various sources. The* E-Cat HT, *which would be off the scale here, lies outside the region occupied by conventional sources.*

**Remarks on the test**
The device subject to testing was powered by 360 W for a total of 96 hours, and produced in all 2034 W thermal. This value was reached by calculating the power transferred by the ECat HT to the environment by convection and power irradiated by the device. The resultant values of generated power density (7093 W/kg) and thermal energy density ($6.81 \cdot 10^5$ Wh/kg) allow us to place the E-Cat HT above conventional power sources.

The procedures followed in order to obtain these results were extremely conservative, in all phases, beginning from the weight attributed to the powder charge, to which the weight of the two metal caps used to seal the container cylinder was added. The same may be said for the choice of attributing an emissivity of 1 to the E-Cat HT; other instances of underestimation may be found in the calculation of the radiating area of the device without the two bases, and in the fact that some parts of the radiating surfaces were covered by metal struts. It is therefore reasonable to assume that the thermal power released by the device during the trial was higher than the values given by our calculations.

Lastly, it should be noted that the device was deliberately shut down after 96 hours of operation. Therefore, from this standpoint as well, the energy obtained is to be considered a lower limit of the total energy which might be obtained over a longer runtime.

This test enabled us to pinpoint several procedural issues, first of all the fact that the device was already in operation when the trial began. This prevented us from correctly weighing the device beforehand, and conducting a thermal analysis of the same without the powder charge, prior to evaluating its yield with the charge in position. The choice of placing the thermal camera under the *E-Cat HT* should also be considered unsatisfactory, as was the impossibility of evaluating the real emissivity of the cylinder's paint coating.

All these issues were taken notice of in the light of the subsequent test held in March. This was performed with a device of new design, as a result of technological improvements effected by Leonardo Corporation in the intervening months.

# PART 2: THE MARCH TEST

**Device and experimental set-up**

The March test was performed with a subsequent prototype of the *E-Cat HT*, henceforth termed *E-Cat HT2*, differing from the one used in the December test both in structure and control system. Externally, the device appears as a steel cylinder, 9 cm in diameter, and 33 cm in length, with a steel circular flange at one end 20 cm in diameter and 1 cm thick. The only purpose of the flange was to allow the cylinder to be inserted in one of various heat exchangers, which are currently under design. As in the case of the previous model, here too the powder charge is contained within a smaller AISI 310 steel cylinder (3 cm in diameter and 33 cm in length), housed within the *E-Cat HT2* outer cylinder together with the resistor coils, and closed at each end by two AISI 316 steel caps.

The outer surface of the *E-Cat HT2* and one side of the flange are coated with black paint, different from that used for the previous test. The emissivity of this coating, a Macota® enamel paint capable of withstanding temperatures up to 800°C, is not known; moreover, it was not sprayed uniformly on the device, as may be seen from the non-uniform distribution of colors in adjacent areas in the thermal imaging.

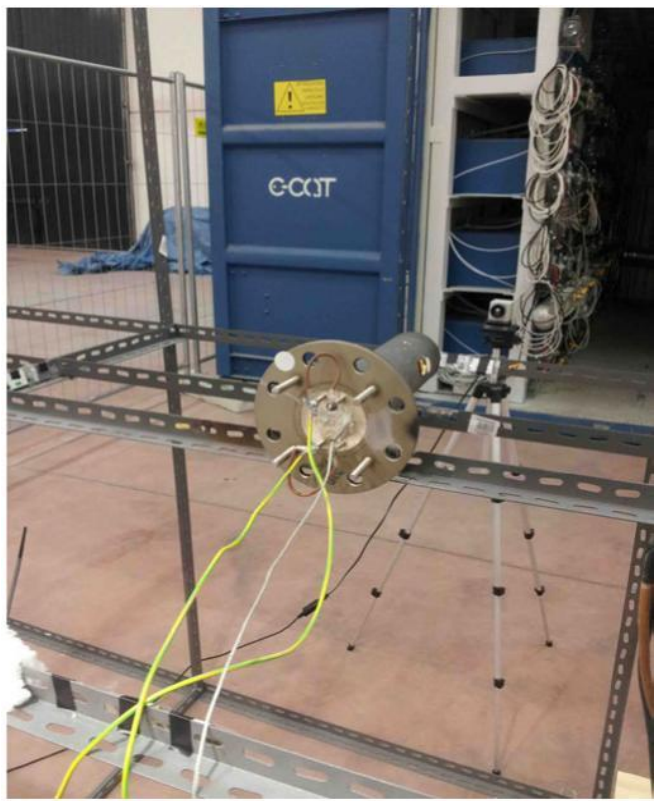

*Fig. 10. Flange of the* E-Cat HT2 *used for the March test. The flange is meant to facilitate insertion of the device in a heat exchanger. Electrical power is fed through the two yellow wires. The third connection was verified to be a PT100 sensor, used to give a feedback temperature signal to the control box in order to regulate the ON/OFF cycle.*

The *E-Cat HT2*'s power supply departs from that of the device used in December in that it is no longer three-phase, but single-phase: the TRIAC power supply has been replaced by a control circuit having three-phase power input and single-phase output, mounted within a box, the contents of which were not available for inspection, inasmuch as they are part of the industrial trade secret. But the main difference between the *E-Cat HT2* and the previous model lies in the control system, which allows the device to work in self-sustaining mode, i.e. to remain operative and active, while powered off, for much longer periods of time with respect to those during which power is switched on. During the test experiment we observed that, after an initial phase lasting about two hours, in which power fed to the resistor coils was gradually increased up to operating regime, an ON/OFF phase was reached, automatically regulated by the temperature feedback signal from a PT100 sensor.

In the ON/OFF phase, the resistor coils were powered up and powered down by the control system at observed regular intervals of about two minutes for the ON state and four minutes for the OFF state. This operating mode was kept more or less unchanged for all the remaining hours of the test. During the OFF state, it was possible to observe – by means of the video displays connected to the IR cameras (see below) – that the temperature of the device continued to rise for a limited amount of time. The relevant data for this phenomenon are displayed in the final part of the present text.

The instrumentation used for the experiment was the same as that of the previous test, with the sole addition of another IR camera, used to measure the temperature of the base (henceforth: "breech") of the *E-Cat HT2*, and of the flange opposite to it. The second camera was also an Optris PI 160 Thermal Imager, but mounting 48° × 37° lens. Both cameras were mounted on tripods during data capture, with the *E-Cat HT2* resting on metal struts. This made it possible to solve two of the issues experienced during the December test, namely the lack of information on the *E-Cat HT2* breech, and the presence of shadows from the struts in the IR imagery.

As in the previous test, the LCD display of the electrical power meter (PCE-830) was continually filmed by a video camera. The clamp ammeters were connected upstream from the control box to ensure the trustworthiness of the measurements performed, and to produce a non-falsifiable document (the video recording) of the measurements themselves.

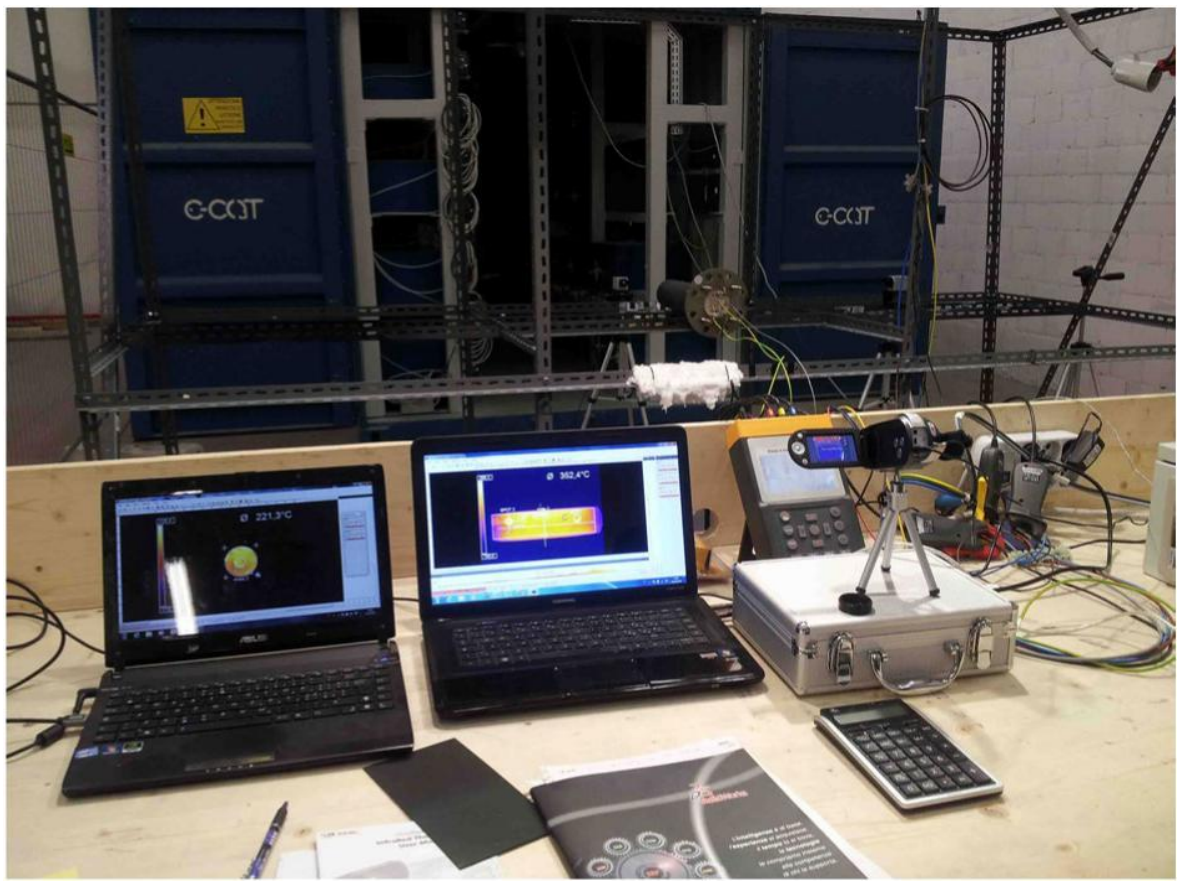

*Fig. 11. Instrument set-up for the test. From left to right: the two laptops connected to the IR cameras, framing the breech (i.e. the base opposite the flange) of the* E-Cat HT2 *and one of its sides, respectively, plus the video camera, and the PCE-830. Background: the* E-Cat HT2 *resting on metal struts and the two IR cameras on tripods.*

Another critical issue of the December test that was dealt with in this trial is the evaluation of the emissivity of the *E-Cat HT2*'s coat of paint. For this purpose, self-adhesive samples were used:

white disks of approximately 2 cm in diameter (henceforth: dots) having a known emissivity of 0.95, provided by the same firm that manufactures the IR cameras (Optris part: ACLSED). According to the manufacturer, the maximum temperature tolerated by a dot before it is destroyed is approximately 380°C. In the course of the test, numerous dots were applied along the side and the breech of the *E-Cat HT2*, but the ones applied to the more central areas showed a tendency to fall off, and had to be periodically replaced. Actually, the distribution of temperatures along the device is non-uniform, and the central part of the cylindrical body is where the temperature reaches values closest to the uppermost working limit for the dots themselves.

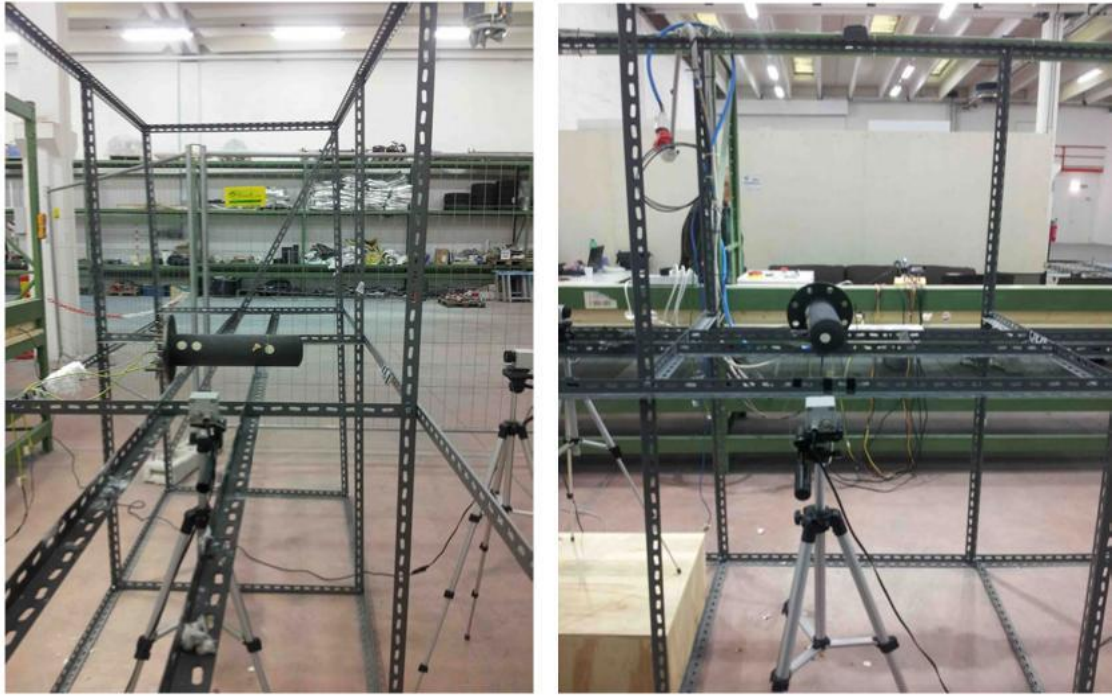

*Fig. 12. Two images of the* E-Cat HT2. *The first displays one of the sides, as framed by one thermal camera, the second displays the breech, framed by the second IR camera. The adhesive "dots" used to evaluate paint emissivity are visible in both images. Note how, in the image on the left, one of the dots is about to fall off due to the heat from the underlying area.*

The dots allow one to determine the emissivity of the surface they are applied to, by comparing the temperatures recorded on the dots and those of the adjacent areas. This procedure may also be applied during data analysis, directly on the completed thermal imagery video. It is possible to divide the thermal images into separate areas – in the same manner as the one used to determine the average temperature of the *E-Cat HT* in the December test – and to assign a specific emissivity to each area. This option proved quite useful later, when analyzing the imagery captured by the cameras, because it made it possible to correct the values of ε that had been assigned during the initial calibration performed while the test was in progress. The dots in the images enabled us to determine that different areas of the device had different emissivity because the paint had not been uniformly applied. Furthermore, it was possible to see how emissivity for each area varied in the course of time, probably on account of a change in the properties of the paint when subjected to continuous heat. For this reason, when analyzing the data after the test, a good number of time intervals were taken into consideration. The thermal images of the *E-Cat HT2* were then divided into areas, and adjusted to the appropriate values of emissivity relevant to every time interval. In order to calculate emitted energy, the temperature then assigned to each area of the device was determined from an average of the various results that had been obtained.

Another improvement over the December test lies in the fact that we were able to perform further measurements (falling outside the 116 hours of the trial run) on the same *E-Cat HT2* used for the test, after removing the inner charge. With this device, termed "dummy", we were able to verify the effectiveness of the methodology used to evaluate the active device, and to estimate the energy emissions of the flange, which would have been difficult to evaluate otherwise.
Lastly, as in the December test, the *E-Cat HT2* was assessed all throughout the test for potential radioactive emissions. The measurements and their analysis were performed once again by David Bianchini, whose report and relevant results are available on demand. His conclusions are quoted below:

*"The measurements performed did not detect any significant differences in exposure and CPM (Counts per Minute), with respect to instrument and ambient background, which may be imputed to the operation of the E-Cat prototypes"*.

**Analysis of data obtained with the "dummy"**

By "dummy" is meant here the same *E-Cat HT2* used for the test described in Part 2, but provided with an inner cylinder lacking both the steel caps and the powder charge. This "unloaded" device was subject to measurements performed after the 116-hr trial run, and was kept running for about six hours. Instrumentation and data analysis were the same as those used for the test of the active *E-Cat HT2*. We prefer to present the data relevant to the dummy beforehand, since these data made it possible to perform a sort of "calibration" of the *E-Cat HT2*, as shall be pointed out below.

The electrical power to the dummy was handled by the same control box, but without the ON/OFF cycle of the resistor coils. Thus, the power applied to the dummy was continuous.

Power to the dummy's resistor coils was stepped up gradually, waiting for the device to reach thermal equilibrium at each step. In the final part of the test, the combined power to the dummy + control box was around 910-920 W. Resistor coil power consumption was measured by placing the instrument in single-phase directly on the coil input cables, and was found to be, on average, about 810 W. From this one derives that the power consumption of the control box was approximately = 110-120 W. At this power, the heat produced from the resistor coils alone determined an average surface temperature (flange and breech excluded) of almost 300°C, very close to the average one found in the same areas of the *E-Cat HT2* during the live test.

Various dots were applied to the dummy as well. A K-type thermocouple heat probe was placed under one of the dots, to monitor temperature trends in a fixed point. The same probe had also been used with the *E-Cat HT2* to double check the IR camera readings during the cooling phase. The values measured by the heat probe were always higher than those indicated by the IR camera: this difference, minimal in the case of the *E-Cat HT2*, was more noticeable in the dummy, where temperature readings proved to be always higher by about 2°C. The most likely reason for the difference is to be sought in the fact that the probe, when covered with the dot securing it the surface, could not dissipate any heat by convection, unlike the areas adjacent to it.

In order to evaluate the power emitted by the dummy by radiation and convection, we decided to divide the image of the cylindrical body into 5 areas, to each of which, by means of dots, we assigned an average emissivity of 0.80. Lastly, the analysis of images relevant to the breech determined for this area another value for ε: 0.88.

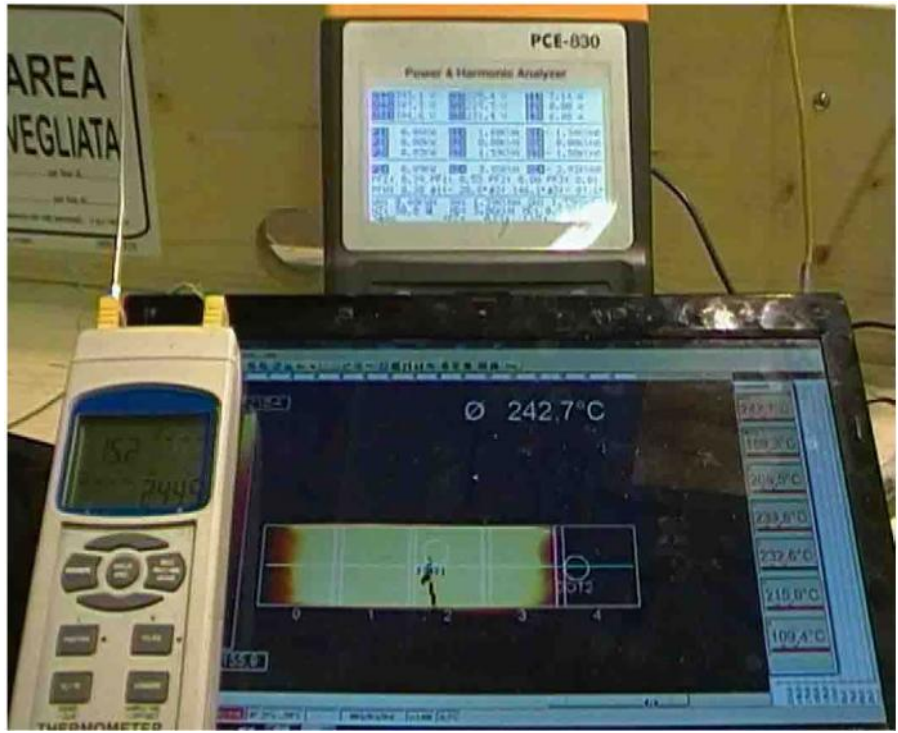

*Fig. 13. Dummy measurement set-up. Center: laptop display showing the thermal image of the dummy divided into 5 areas, and the dark shadow of the thermocouple, with probe point located under a dot. Left: thermocouple LCD display, indicating a temperature of 244.5°C. This is relevant to the same area which the IR camera reading of 242.7°C, visible on the laptop display, refers to. The difference is most likely caused by lower thermal exchange between the probe and the environment..*

For each of the five areas, energy emitted by radiation was calculated. Once again, Stefan-Boltzmann's formula multiplied by the area taken into consideration was used, as in Part 1, equation (5). Power emitted by convection was calculated by equations (9) and (10). The equations are repeated below for clarity's sake, followed by a table summarizing the results.

$E = \varepsilon \sigma A T^4$ [W]  (5)

$Q = hA(T-T_f) = hA\Delta T$ [W]  (9)

$h = C'' (T-T_f)^n D^{3n-1}$  (10)

$Area_{Dummy} = 2\pi RL = 989.6 \cdot 10^{-4}$ [m$^2$]

$Area_{Top} = \pi R^2 = 63.61 \cdot 10^{-4}$ [m$^2$]

Note that coefficients C'' and *n* of (10) have the same value calculated for the December test, namely C'' = 1.32, and *n* = 0.25, whereas the diameter D is now = 9 cm.
Moreover, $Area_{Dummy}$ refers to the cylindrical body of the device without flange or breech.
Lastly, the contributing factor due to ambient temperature, termed "*E*_room" in (7) above, has already been subtracted from the power values associated with each area. This was calculated assuming an ambient temperature value of 14.8°C.

$E\_room = (5.67 \cdot 10^{-8})(288)^4 (0.80)(198 \cdot 10^{-4}) = 6.18$ [W]

|        | **Area 1** | **Area 2** | **Area 3** | **Area 4** | **Area 5** | **Sum** |
|--------|------------|------------|------------|------------|------------|---------|
| E (W)  | 84.9       | 112        | 109        | 102        | 49.3       | 457.2   |
| Q (W)  | 53.7       | 63         | 62.6       | 59.9       | 38.3       | 277.5   |
| W Total| 138.6      | 175        | 171.6      | 161.9      | 87.6       | 734.7   |

*Table 6. Power emitted by radiation (E) and convection (Q) for each of the five areas. The value of* E_room, *about 6.18 W, has already been subtracted from power* E *in the relevant area.*

Using the second thermal imagery camera, it was possible to monitor the temperature of the breech, which was almost stable at 225°C. We were thus able to compute the contributing factor to the total radiating energy associated with this part of the dummy: a value of $E-E\_room = 17$ W.
As for the flange, it was not possible to evaluate its temperature with sufficient reliability, despite the fact that it was partially framed by both IR cameras. A careful analysis of the relevant thermal imagery revealed how part of the heat emitted from the flange was actually reflected heat coming from the body of the dummy. In fact, the position of the flange is such that one of its sides constantly receives radiative heat emitted by the body of the cylinder: if we were to attribute the recorded temperature to the flange, we would risk overestimating the total radiative power.
Conservation of energy was used to evaluate the contributing factor of the flange, and of all other not previously accounted factors, to the total energy of the dummy. Thus, we get:

$$810 \,[W] - (735 + 17)\,[W] = 58\,[W] \tag{22}$$

This last value is the sum of the contributive factors relevant to all unknown values, namely: flange convection and radiation, breech convection (NB convection only), losses through conduction, and the margin of error associated with our evaluation.
Since the temperatures reached by the dummy and by the *E-Cat HT2* during their operation were seen to be quite similar, this value will also be used to calculate the power relevant to the *E-Cat HT2*, where it will be attributed the same meaning.

**Analysis of data obtained with the *E-Cat HT2***
The *E-Cat HT2* was started approximately at 3:00 p.m. on March 18. The initial power input was about 120 W, gradually stepping up during the following two hours, until a value suitable for triggering the self-sustaining mode was reached. From then onwards, and for the following 114 hours, input power was no longer manually adjusted, and the ON/OFF cycles of the resistor coils followed one another at almost constant time intervals. During the coil ON states, the instantaneous power absorbed by the *E-Cat HT2* and the control box together was visible on the PCE-830 LCD display. This value, with some fluctuations in time, remained in any case within a range of 910-930 W. The PCE-830 LCD display showed the length of the ON/OFF intervals: with reference to the entire duration of the test, the resistor coils were on for about 35% of the time, and off for the remaining 65%.
As in the case of the dummy, in order to determine the average temperatures for the *E-Cat HT2* we opted to divide its thermal images into five areas, plus another one for the breech. An analysis of various time segments (about five hours each), taken in the course of each day of the test, revealed that the behavior of the device remained more or less constant, and became quite stable especially from the third day onwards. Using the same procedure as before, we obtained an average temperature for each of the five areas, thereafter employing equations (5), (9), and (10) in order to calculate power emitted by radiation and convection, respectively.

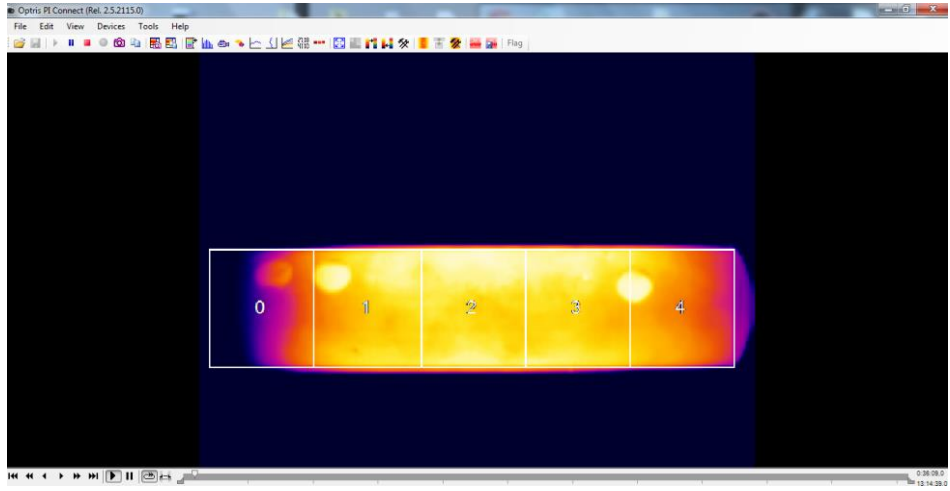

*Fig. 14. 5-area division of the* E-Cat HT2 *image. The flange does not appear in the image because the display range chosen for the IR camera does not detect objects colder than 150°C.*

Emissivity values for each area were adjusted in each IR camera video sample thanks to the continuing presence of dots: according to position and time, the found values for ε fluctuated between a low of 0.76 and a high of 0.80. Areas subject to the most intense heat were seen to have slightly higher emissivity compared to peripheral ones, and all showed a slight upward trend as the test progressed, probably because of a change in the properties of the paint.

In order to account for a certain degree of arbitrariness inherent in this method of evaluation, it was decided to assign a reference temperature to the various areas into which the *E-Cat HT2* had been divided. This was obtained by assigning to all areas the most frequently found value for ε and associating a percentage error to it. This error is the result of the difference between two extreme values, namely the temperature obtained by assigning to all areas the lowest level of emissivity ever found in any one of them (= 0.76), and the temperature obtained assigning to all areas the highest value for ε ever found (= 0.80). Tables 7 and 8 summarize the results: the first refers to the average of temperatures in each of the five areas for different values of ε, whereas the second gives the average values of power emitted by radiation (*E*) and convection (Q) for different values of ε, while taking into account the sum performed on the five areas.

| ε | T 1 (°C) | T 2 (°C) | T 3 (°C) | T 4 (°C) | T 5 (°C) | Average |
|---|---|---|---|---|---|---|
| Average ε | 261.0 | 319.4 | 326.0 | 318.3 | 286.9 | 302.3 |
| 0.76 | 261.0 | 328.4 | 335.2 | 327.3 | 286.9 | 307.7 |
| 0.80 | 254.0 | 319.4 | 326.0 | 318.4 | 279.2 | 299.4 |

*Table 7. Average temperatures relevant to the divisions into five areas of the* E-Cat HT2*'s cylindrical body, calculated according to average values of emissivity (first row), absolute minimal values (second row), and absolute maximum values (third row), collated by taking into consideration all the areas and all the analyzed time intervals. The last column gives the averages of the previous values for each of the five areas.*

| E | E (W) | Q(W) | E(W) + Q(W) |
|---|---|---|---|
| Average ε | 459.8 | 281.5 | 741.3 |
| 0.76 | 463.8 | 288.2 | 752.0 |
| 0.80 | 458.6 | 277.9 | 736.6 |

*Table 8. Emitted power values by radiation (*E*) and by convection (Q) for different values of ε. The numbers are computed from the power average of all five areas, minus the* E_room *component arising from the contributing factor of ambient temperature.*

The error associable to the average value of emitted power may be got by taking into account the difference between what is obtained by attributing to each area the highest possible and the lowest possible value for ε. Thus:

$$(752.0-736.6)/741.3 = 2\% \quad (23)$$

As may be inferred from the last value above, the uncertainty regarding emissivity does not affect the results much, and should therefore be considered a parameter of lesser critical import than what was originally estimated.

The average temperature relevant to the breech, as well as its average emissivity, turned out to be extremely constant over time, with values of 224.8°C and 0.88, respectively. We can therefore associate them with a value of irradiated power $E-E\_room = 17$ [W].

At this point, all the contributing factors relevant to the thermal power of the *E-Cat HT2* are available, i.e. the power emitted by the cylindrical body through radiation and convection, the power emitted by radiation by the breech, and the set of missing factors (conduction, breech convection, flange radiation and convection). It is now possible to obtain a complete estimate:

$$Emitted\ Power_{E\text{-}Cat\ HT2} = (741.3 + 17 + 58)\ [W] = (816.3 \pm 2\%)\ [W] = (816 \pm 16)\ [W] \quad (24)$$

**Ragone Chart**

Upon completion of the test, the *E-Cat HT2* was opened, and the innermost cylinder, sealed by caps and containing the powder charges, was extracted. It was then weighed (1537.6 g) and subsequently cut open in the middle on a lathe. Before removal of the powder charges, the cylinder was weighed once again (1522.9 g), to compensate for the steel machine shavings lost. Lastly, the inner powders were extracted by the manufacturer (in separate premises we did not have access to), and the empty cylinder was weighed once again (1522.6 g). The weight that may be assigned to the powder charges is therefore on the order of 0.3 g; here it shall be conservatively assumed to have value of 1 g, in order to take into account any possible source of error linked to the measurement.

According to the data available from the PCE-830 analyzer, the overall power consumption of the *E-Cat HT2* and the control box combined was 37.58 kWh. The associated instantaneous power varied between 910 and 930 W during the test, so it may be averaged at 920±10 W. In order to determine the power consumption of the *E-Cat HT2* alone, one must subtract from this value the contributive factor of the control box power consumption. As it was not possible to measure the latter while the test on the *E-Cat HT2* was in progress, one may refer to the power consumption of the box measured during the dummy test. This value would in all likelihood be higher in the case of operative *E-Cat HT2*, due to the electronic circuits controlling the self-sustaining mode: so, as usual, we shall adopt the more conservative parameter.

If one assumes that the control box absorbed about 110 W, we can associate the *E-Cat HT2* with a consumption of:

$$Instantaneous\ Power\ Consumption_{E\text{-}Cat\ HT2} = (920 - 110)\ [W] = 810\ [W] \quad (25)$$

Keeping in mind the fact that this consumption was not constant over time, but may be referred just to 35% of the total test hours, one may calculate the effective power consumption of the device as:

$$Effective\ Power\ Consumption_{E\text{-}Cat\ HT2} = (810/100) \cdot 35 = 283.5\ [W] \quad (26)$$

Let us further assume an error of 10%, in order to include any possible unknown source. Errors of this extent are commonly accepted in calorimetric measurements, and in our case they would comprise various sources of uncertainty: those relevant to the consumption measurements of the *E-Cat HT2* and the control box, those inherent in the limited range of frequencies upon which the IR cameras operate, and those linked to the calculation of average temperatures.

The energy produced by the *E-Cat HT2* during the 116 hours of the test is then:

*Produced Energy*$_{E\text{-}Cat\ HT2}$ = (816-283.5) · 116 = (6.2 ± 0.6) · $10^4$ [Wh] (27)

From (27) one may gather the parameters necessary to evaluate the position held by the *ECat HT2* with respects to the Ragone Plot, where specific energy is represented as a function on a logarithmic scale of the specific power of the various energy storage technologies [see Ref. 8].

For power density we have:

(816-283.5)/0.001 = 532500 [W/kg] ~ 5 · $10^5$ [W/kg] (28)

Thermal energy density is obtained by multiplying (28) by the number of test hours:

532500 · 116 = (6.2 ± 0.6) · $10^7$ [Wh/kg] ~ 6 · $10^7$ [Wh/kg] (29)

It is easy to infer from the Ragone chart, another example of which may be seen below in fig. 15 below, that these values place the *E-Cat HT2* at about three orders of magnitude beyond any other conventional chemical energy source.

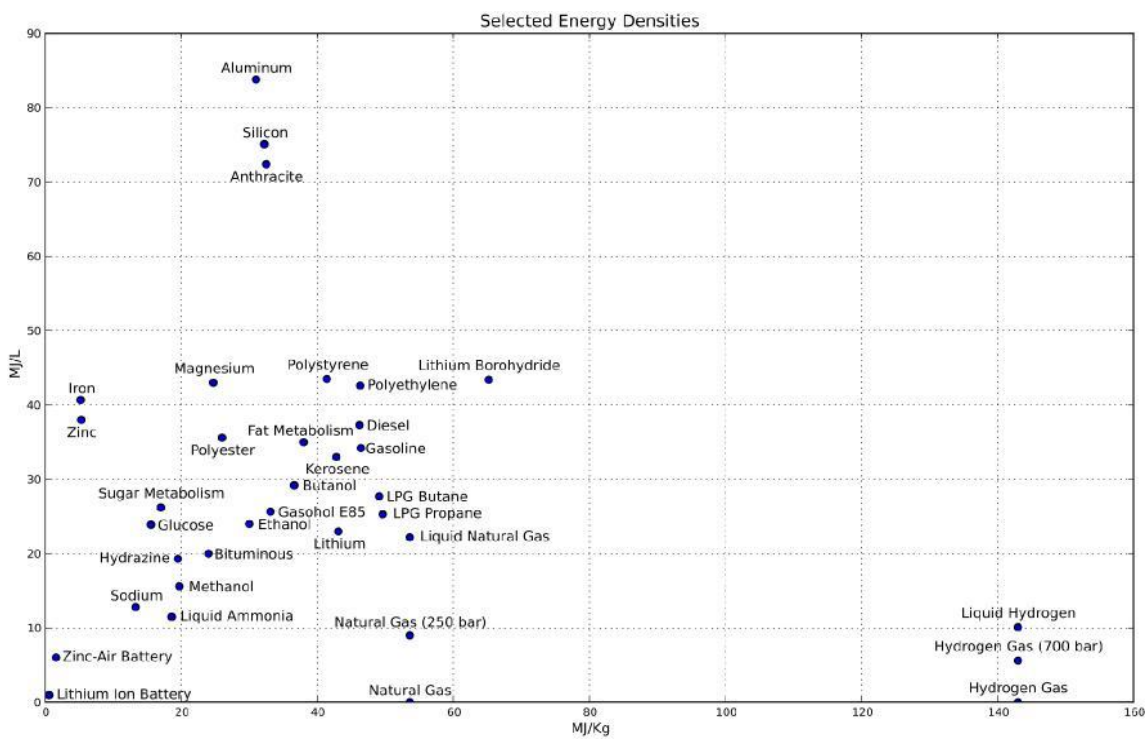

*Fig. 15. Another version of the Ragone Plot of Energy Storage [Ref. 8]. In this plot, specific volumetric and gravimetric energy densities are presented for various sources. The ECatHT2, out of scale here, lies outside the region occupied by conventional chemical sources.*

As it was not possible to inspect the inside of the control box, let us now repeat the last calculations supposing, as a precautionary measure, that all power consumption is assigned to the *E-Cat HT2*.

According to this logic, and assigning to the *E-Cat HT2* the maximum value of error given by (24), namely (816 - 16)W = 800 W, one gets:

*Conservative Power Consumption*$_{E\text{-}Cat\ HT2}$ = (920/100) · 35 = (322 ± 32) [W] (30)

whereas (28) and (29) become:

$(800-322)/0.001 = (4.7 \pm 0.5) \cdot 10^5$ [W/kg] (31)

$478000 \cdot 116 = (5.5 \pm 0.6) \cdot 10^7$ [Wh/kg] (32)

The results thus obtained are still amply sufficient to rule out the possibility that the *E-Cat HT2* is a conventional source of energy.

Let us associate to this last value of conservative power consumption the worst-case scenario:

$(322 + 32)$ [W] = 354 [W] (33)

Then the values of power density and energy density would then be:

$(800-354)/0.001 = (4.4 \pm 0.4) \cdot 10^5$ [W/kg] (34)

$446000 \cdot 116 = (5.1 \pm 0.5) \cdot 10^7$ [Wh/kg] (35)

Obviously, not even in this case do we have any substantial change as far as the position occupied by the *E-Cat HT2* in the Ragone plot is concerned.

For a further confirmation of the fact that the *E-Cat HT2*'s performance lies outside the known region of chemical energy densities, one can also calculate the volumetric energy density of the reactor, by referring to the whole volume occupied by the internal cylinder, namely $1.5^2 \cdot \pi \cdot 33 = 233$ cm$^3$ = 0.233 l. This is the most conservative and "blind" approach possible.

Taking the figures from the worst case, we get a net power of 800-354=446 W; by multiplying this by $(3600 \cdot 116)$, we find that 185 MJ where produced. Thus, we have a volumetric energy density of $185/0.233 = (7.93 \pm 0.8)10^2$ MJ/Liter, meaning that even by resorting to the most conservative and "worst case scenarios", where the total volume of the reactor is comprehensive of the 5-mm thick steel cylinder, we see that we are still at least one order of magnitude above the volumetric energy density of any known chemical source [Ref. 8].

### *E-Cat HT2* performance calculation

According to the engineering definition, COP is given by the ratio between the output power of a device and the power required by its operation, thereby including, in our case, the power consumed by the control electronics.

For the *E-Cat HT2* one would therefore have (assuming a 10% uncertainty in the powers):

COP = 816/322 = 2.6 ± 0.5 (36)

In order to compare this figure with the COP value obtained in the December test (5.6; see (19)), one must first of all consider that the two values were obtained in different experimental contexts: (19) gives the ratio between power emitted and power consumed by the *E-Cat HT* only, without the TRIAC power supply, whereas (36) includes power consumption by the *E-Cat HT2*'s control device instrumentation. The expression useful for such a comparison is therefore the following:

COP = 816/283 = 2.9 ± 0.3 (37)

Thus, (19) and (37) give the performances specific to prototypes *E-Cat HT* and *E-Cat HT2*, respectively – regardless of the electronic circuits (also prototypes) used to control them. Since the main goal of the present paper is a specific investigation of E-Cats as physical systems, these are the most meaningful expressions for our purposes.

The reasons for the appreciable difference between the value obtained in December and the one found in March are probably to be sought in the tendency of the COP to increase with temperature, a fact which was noticed even in the November test. In that occasion, reaching a certain critical temperature threshold was enough to cause the reaction to diverge uncontrollably and destroy the device. Considering that, in December, the *E-Cat HT*'s average temperature was 438°C, vs an average of 302°C for the *E-Cat HT2* in March, a higher COP for the former device with respect to that found in the latter was by no means unexpected.

In any event, inasmuch as the quantity of the charge contained in the first device is not known, a comparison between the two tests is not strictly appropriate. It is possible that the two coefficients of performance differ only because the quantity of powder used in the two tests was different.

**Remarks on the test**

An interesting aspect of the *E-Cat HT2* is certainly its capacity to operate in self-sustaining mode. The values of temperature and production of energy which were obtained are the result of averages not merely gained through data capture performed at different times; they are also relevant to the resistor coils' ON/OFF cycle itself. By plotting the average temperature vs time for a few minutes of test (Plot 3) one can clearly see how it varies between a maximum and a minimum value with a fixed periodicity.

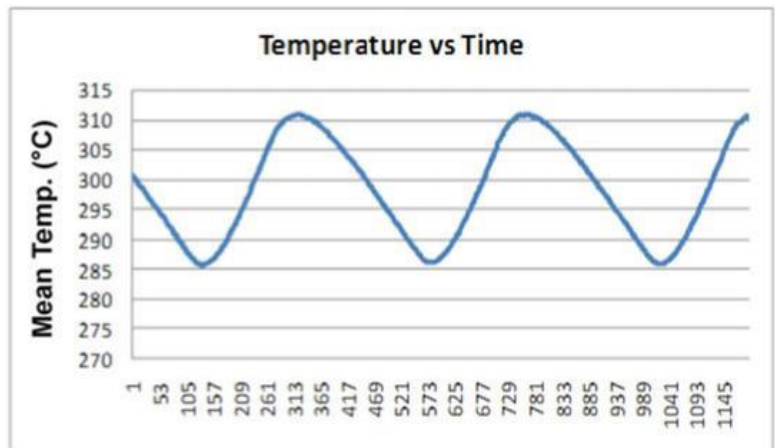

*Plot 3. Average surface temperature trend of the E-Cat HT2 over several minutes of operation. Note the heating and cooling trends of the device, which appear to be different from the exponential characteristics of generic resistor.*

Looking at Plot 3, the first feature one notices is the appearance taken by the curve in both the heating and cooling phases of the device. If we compare these in detail with the standard curves of a generic resistor (Plot 4 and Plot 5), we see that the former differ from the latter in that they are not exponential .

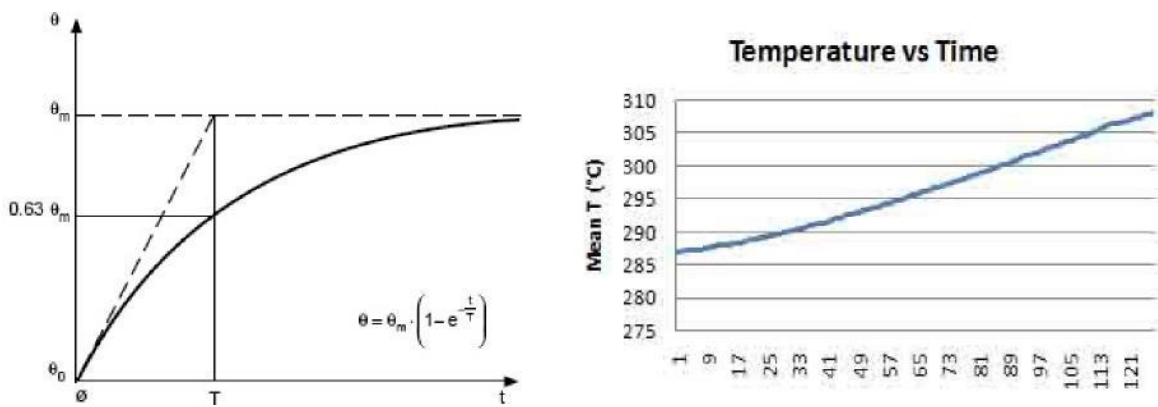

*Plot 4. Comparing the typical heating curve of a generic resistor (left, [Ref. 9]) to the one relevant to one of the E-Cat HT2's ON states.*

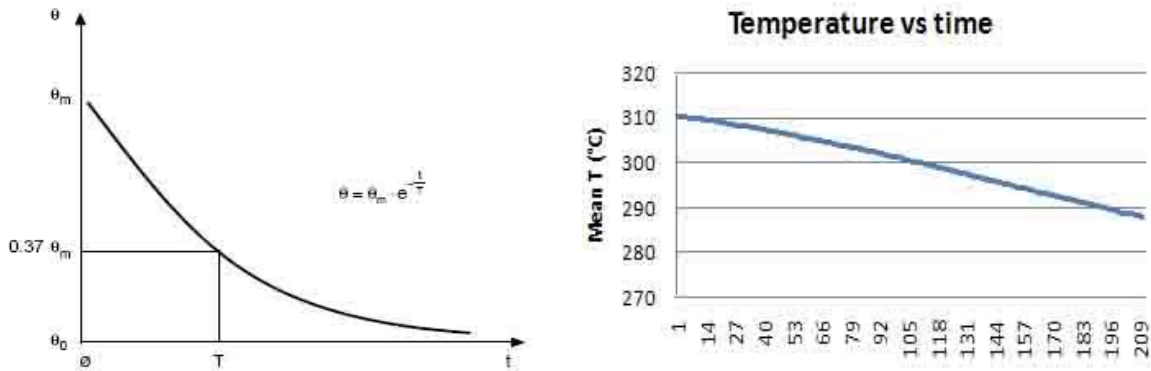

*Plot 5. Comparing the typical cooling curve of a generic resistor (left, [Ref. 9]) to the one relevant to one of the E-Cat HT2's OFF states.*

Finally, the complete ON/OFF cycle of the *E-Cat HT2*, as seen in Plot 3, may be compared with the typical heating-cooling cycle of a resistor, as displayed in Plot 6.

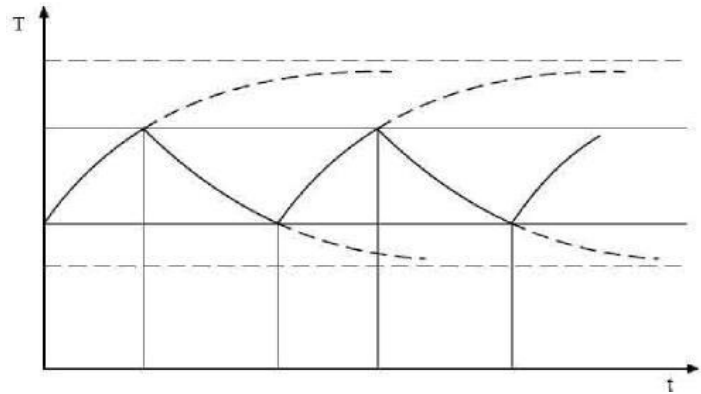

*Plot 6. Heating and cooling cycle of a generic resistor [Ref. 9]. The trend is described by exponential type equations.*

What appears indicated here is that the priming mechanism pertaining to some sort of reaction inside the device speeds up the rise in temperature, and keeps the temperatures higher during the cooling phase.
Another very interesting behavior is brought out by synchronically comparing another set of curves: power produced over time by the *E-Cat HT2*, and power consumed during the same time. An example of this may be seen in Plot 7, which refers to about three hours of test. The resistor coils ON/OFF cycle is plotted in red, while the power-emission trend of the device appears in blue.

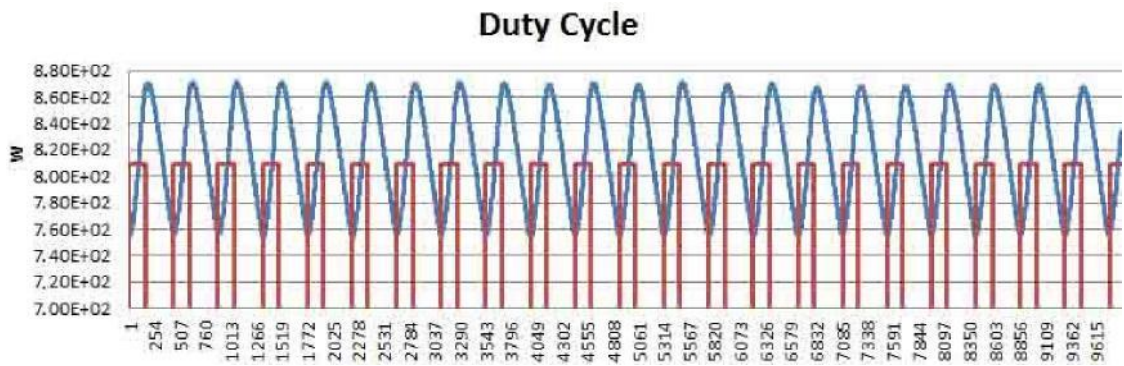

*Plot 7. Chart showing emitted power (in blue) and consumed power (in red) vs time for the E-Cat HT2.*

Starting from any lowest point of the blue curve, one can distinguish three distinctive time intervals. In the first, emitted power rises, while remaining below the red line representing consumed power. In the second, emitted power rises above consumed power, and approaches its peak while the resistors are still on. In the third, after the resistors have been turned off, emitted power reaches its peak and then begins to fall to a new minimum value, whereupon the resistors turn on again. In the first time interval, emitted power is less than consumed power; but already in the second the trend reverses, and continues as such into the beginning of the third. Plot 8, which gives an expanded view of Plot 7, the three intervals are visually enhanced for the sake of clarity.

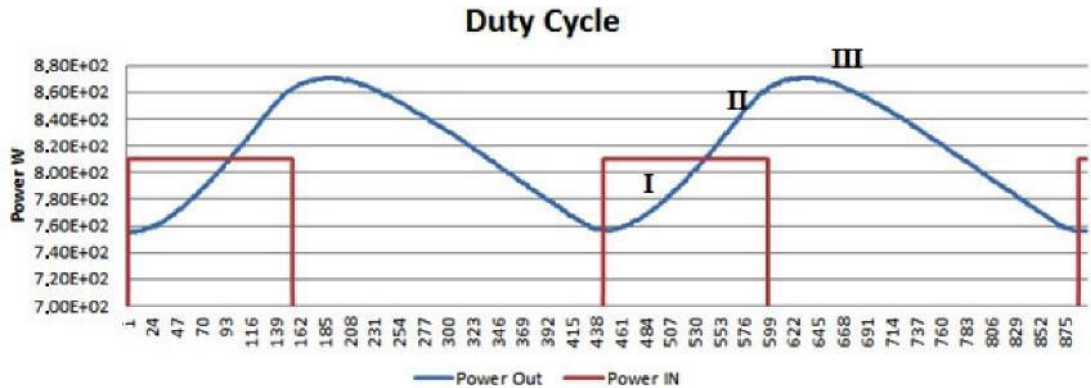

*Plot 8. Detail taken from Plot 7, reproducing the first two periods of the cycle. The three time intervals in which each period may be divided are labeled by Roman numerals.*

Further food for thought may be found by analyzing the trend of the ratio between energy produced and energy consumed by the *E-Cat HT2*, during the time interval shown in Plot 7. The blue curve in Plot 9 is the result of the analysis, and is reproduced here together with the red curve of power consumption normalized to 1. Basically, for every second taken into account, the corresponding value of the blue curve is calculated as the ratio between the sum of the power per second emitted in all the previous seconds, and the sum of the power per second consumed in all the previous seconds.

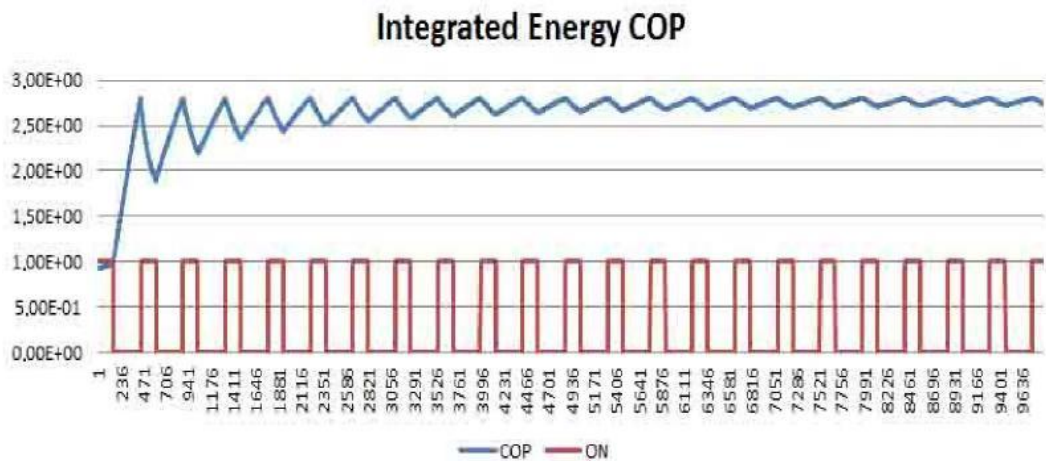

*Plot 9. The blue curve is the result of the ratio between energy produced and consumed by the E-Cat HT2, with reference to the same time instants dealt with in Plot 7. The red curve represents the ON/OFF trend of the resistor coils normalized to 1.*

All the above trends are remarkable, and warrant further in-depth enquiry.

**Conclusions**

The two test measurements described in this text were conducted with the same methodology on two different devices: a first prototype, termed *E-Cat HT*, and a second one, resulting from technological improvements on the first, termed *E-Cat HT2*. Both gave indication of heat production from an unknown reaction primed by heat from resistor coils. The results obtained indicate that energy was produced in decidedly higher quantities than what may be gained from any conventional source. In the March test, about 62 net kWh were produced, with a consumption of about 33 kWh, a power density of about $5.3 \cdot 10^5$, and a density of thermal energy of about $6.1 \cdot 10^7$ Wh/kg. In the December test, about 160 net kWh were produced, with a consumption of 35 kWh, a power density of about $7 \cdot 10^3$ W/kg and a thermal energy density of about $6.8 \cdot 10^5$ Wh/kg. The difference in results between the two tests may be seen in the overestimation of the weight of the charge in the first test (which included the weight of the two metal caps sealing the cylinder), and in the manufacturer's choice of keeping temperatures under control in the second experiment to enhance the stability of the operating cycle. In any event, the results obtained place both devices several orders of magnitude outside the bounds of the Ragone plot region for chemical sources.

Even from the standpoint of a "blind" evaluation of volumetric energy density, if we consider the whole volume of the reactor core and the most conservative figures on energy production, we still get a value of $(7.93 \pm 0.8) \, 10^2$ MJ/Liter that is one order of magnitude higher than any conventional source.

Lastly, it must be remarked that both tests were terminated by a deliberate shutdown of the reactor, not by fuel exhaustion; thus, the energy densities that were measured should be considered as lower limits of real values.

The March test is to be considered an improvement over the one performed in December, in that various problems encountered in the first experiment were addressed and solved in the second one. In the next test experiment which is expected to start in the summer of 2013, and will last about six months, the long term performance of the *E-Cat HT2* will be tested. This test will be crucial for further attempts to unveil the origin of the heat phenomenon observed so far.

**Acknowledgments**

The authors would like to thank David Bianchini, M.Sc. for his cooperation in performing the test. We also wish to thank Prof. Ennio Bonetti (Bologna University), Pierre Clauzon, M.Eng. (CNAM-CEA Paris), Prof. Loris Ferrari (Bologna University), and Laura Patrizii, Ph.D. (INFN) for their helpful discussions, Prof. Alessandro Passi (Bologna University [ret.]) for his patient work in translating the text.

We would especially like to thank Andrea Rossi, M.A., inventor of the E-Cat, for giving us the opportunity to independently test the apparatus, and Prof. Em. Sven Kullander and Prof. Björn Gålnander (Uppsala University) for their continued interest in and support for these investigations.

A special thought and warm thanks must be also expressed to Prof. Em. Sergio Focardi (Bologna University) and Prof. Em. Hidetsugu Ikegami (Osaka University).

The authors would like to express their appreciation to Optris GmbH and Luchsinger Srl for their support and technological assistance.

Financial support from Alba Langenskiöld Foundation and ELFORSK AB, for the Swedish participation in the E-Cat test experiment, is gratefully acknowledged.

**References**

[1] S. Focardi, R. Habel and F. Piantelli, Nuovo Cimento (Brief Notes) 107A (1994), 163.

[2] S. Focardi et al., Nuovo Cimento 111A (1998), 1233.

[3] S. Focardi and A. Rossi, internal report, 2010.

[4] J.M. Coulson and J.F. Richardson, *Chemical Engineering*, 1999 (sixth edition), Butterworth Heinemann.

[5] A. Bejan, A.D. Kraus, *Heat Transfer Handbook*, 2003, John Wiley & Sons Inc.

[6] Optris, *Basic principles of non-contact temperature measurement*, www.optris.com.

[7] Ahmed F. Ghoniem, *Needs, resources and climate change: clean and efficient conversion technologies*, Progress in Energy and Combustion Science 37 (2011), 15-51, fig.38.

[8] http://en.wikipedia.org/wiki/File:Energy_density.svg

[9] Nuova Magrini Galileo, Dossier Tecnico n°9, *Determinazione della sovratemperatura in apparecchi sottoposti a sovracorrenti cicliche*, Merlin Gerin.

**APPENDIX**

This Appendix seeks to shed light on some details regarding the electrical measurements which were performed during the March test.

Figure 1 shows the wiring diagram of the PCE-830.

All cables were checked before measurements began. The ground cable, the presence of which was necessary for safety reasons, was disconnected. The container holding the electronic control circuitry was lying on a wooden plank and was lifted off the surface it was resting on, and checked on all sides to make sure that there were no other connections.

We furthermore made sure that the frame supporting the *E-CAT HT2* was not fastened to the pavement and that there were no cables connected to it.

Therefore, apart from its connections to the control electronics, the *E-CAT HT2* appeared to be electrically insulated.

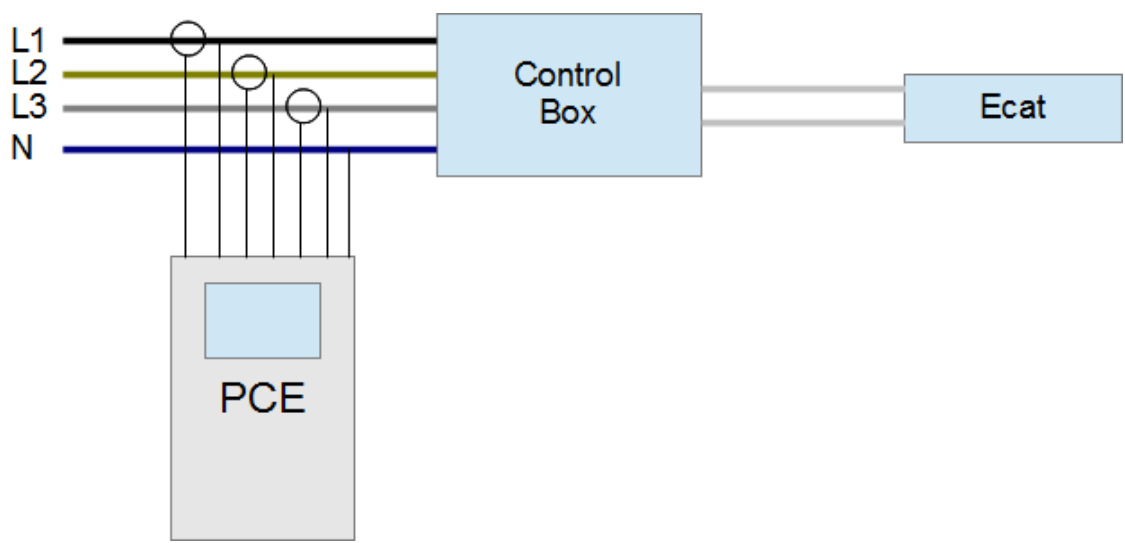

*Fig. 1. Wiring diagram of the PCE-830 Power and Harmonics Analyzer. The three-phase power cables were checked and connected directly to the electrical outlet. It was established and verified that no other cable was present and that all connections were normal. The ground cable was disconnected before measurements began.*

The PCE-830, in addition to providing voltage and current values for each phase, allows one to check both the waveform and its spectral composition in harmonics of the fundamental frequency (50 Hz).

As far as voltage is concerned, the figures, considering that peak values are shown, clearly show that the waveform was sinusoidal and symmetrical, and that there were no levels of DC voltage – having it been already established that there were no other electrical connections.

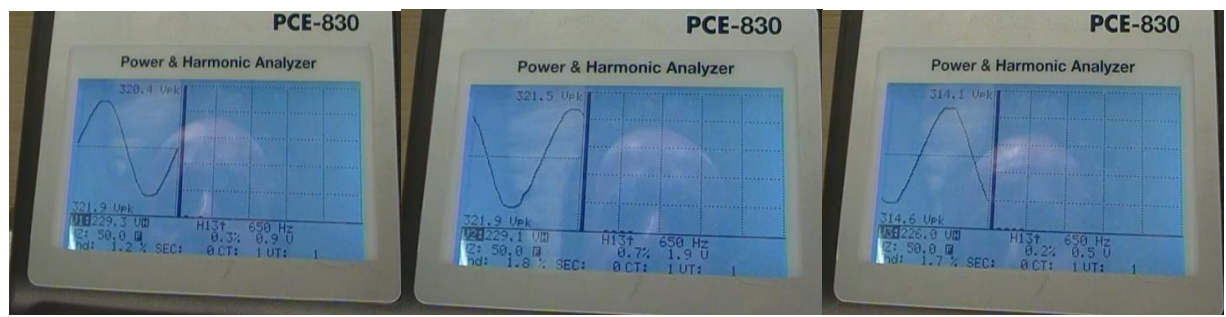

*Fig. 2. Waveforms for voltages measured at control box input. Note the symmetrical sinusoids, with no indication of DC voltage levels. The spectral composition includes only the first harmonic at 50 Hz.*

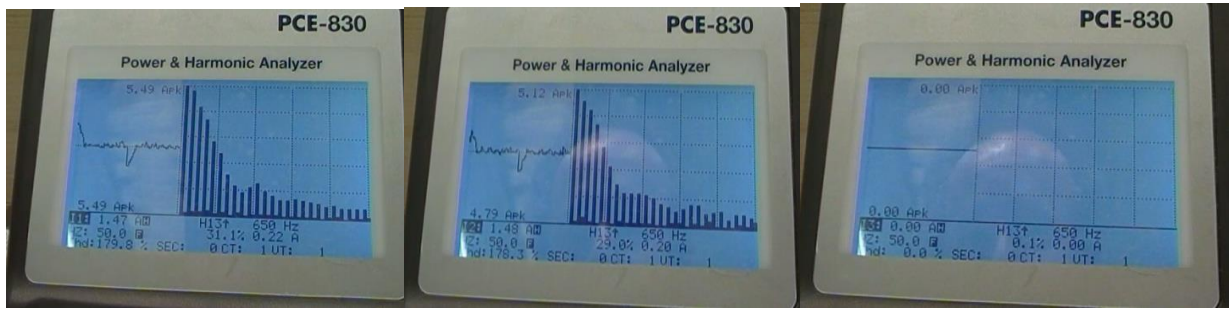

*Fig. 3. Current waveform and harmonics spectrum. The waveform is typical of a sine wave cut by a TRIAC regulator. One may see that no current was present in the third phase. The first 50 harmonics are visualized. Note how the power is for the most part contained within the 30th harmonic.*

The instrument's stated measurement error is 2% within the 20th harmonic, and 5% from harmonics 21 to 50. In our measurements, a margin of error of 10% was assumed.
As far as measurements of current are concerned, it was ascertained that no current was present in the third phase, and that, for the other two phases, the waveform harmonics spectrum, which appeared to be the one normally associated with a TRIAC regulator, was contained within the interval measurable by the instrument.